\documentclass[11pt]{article}
\usepackage{amsmath,amssymb,amsthm}

\usepackage{amsmath}
\usepackage{amssymb}
\usepackage{amsthm}
\usepackage[dvips]{graphicx}
\usepackage{times}
\usepackage{hyperref}
\usepackage[usenames, dvipsnames]{color}
\usepackage{mathrsfs}
\usepackage{enumerate}

\hypersetup{
    unicode=true,           
    pdftoolbar=false,       
    pdfmenubar=false,       
    pdffitwindow=false,     
    pdfstartview={FitH},    
    pdftitle={},            
    pdfauthor={},           
    pdfsubject={},          
    pdfcreator={},          
    pdfproducer={},         
    pdfkeywords={},         
    pdfnewwindow=true,      
    colorlinks=true,        
    linkcolor=ForestGreen,  
    citecolor=blue,         
    filecolor=green,        
    urlcolor=blue           
}

\usepackage[T1]{fontenc}
\usepackage{graphicx}
\usepackage{mathrsfs}
\usepackage{amssymb,amsmath}
\usepackage{amssymb,amsmath}
\newcommand{\beq}[2]{\begin{equation}#1\label{#2}\end{equation}}
\newcommand{\ceq}[1]{(\ref{#1})}

\newfont{\mbld}{cmbx10 scaled 800}
\newfont{\cab}{cmsy10 scaled 1200}
\newfont{\scab}{cmsy10 scaled 1000}
\newfont{\bcall}{cmbsy10 scaled 1200}

\begin{document}
\title{\LARGE\bf Knots, links, anyons and statistical mechanics\\ of entangled polymer rings}
\author{Franco Ferrari$^{\,1}$\footnote{e-mail: franco@feynman.fiz.univ.szczecin.pl}
\hspace{0.5cm}
Jaros{\l}aw Paturej$^{\,1,2}$
\hspace{0.5cm}
Marcin Pi\c{a}tek$^{\,1,3}$\footnote{e-mail: piatek@fermi.fiz.univ.szczecin.pl}
\hspace{0.5cm}
Yani Zhao$^{\,4}$
\\[4pt]
$^{1\,}$Faculty of Mathematics and Physics, University of Szczecin\\
Wielkopolska 15, 70--451 Szczecin, Poland
\\[4pt]
$^{2\,}$Leibniz-Institut f\"{u}r Polymerforschung Dresden e.V.,\\ 01069 Dresden, Germany
\\[4pt]
$^{3\,}$Bogoliubov Laboratory of Theoretical Physics,\\
Joint Institute for Nuclear Research, 141980 Dubna, Russia
\\[4pt]
$^{4\,}$Max Planck Institute for Polymer Research,\\
Ackermannweg 10--D-55128 Mainz, Germany}
\date{}
\maketitle
\vspace{-1cm}
\begin{abstract}
\noindent
The field theory approach to the statistical mechanics of a system of
N polymer rings linked together is generalized to the case of links
that have a fixed number $2s$ of maxima and minima. Such kind of links
are called plats and appear for instance in the DNA of living
organisms. The topological states of the link are distinguished using
the Gauss linking number. This is a relatively weak link invariant in
the case of a general link, but its efficiency improves when
$2s-$plats are considered. It is proved that, if we restrict
ourselves to $2s-$plat conformations, the field theoretical model
established here is able to take into account also the interactions of
topological origin involving three chains simultaneously. It is shown
that these three-body interactions have nonvanishing contributions
when three or more rings are entangled together, enhancing for
instance the attractive forces between monomers. The model can be used
to study the statistical mechanics of polymers in confined geometries,
for instance when $2s$ extrema of a few polymer rings are attached to
membranes. Its partition function is mapped here into that of a
multi-layer electron gas. Such quasi-particle systems are studied in
connection with several interesting applications, including high-$T_c$
superconductivity and topological quantum computing. At the end an
useful connection with the cosh-Gordon equation is shown. 
\end{abstract}

\maketitle

\section{Introduction}
Knots and links are a fascinating subject and
are researched in connection with several concrete applications  both in
physics and biology \cite{grosberg,dna,katritch96,katritch97,krasnow,
laurie,cieplak,marko,liu,wasserman,sumners,vologodski,orlandini,levene,
kurt,mehran,pp2,yan,arsuaga,arsuaga2,
metzler,pieranski,diao,sumners2,faddeev,yeates}. A beautiful review from a
theoretical physics point of view about
knot theory and polymers can be found in Ref.~\cite{kleinert}, Chapter 16.
In this paper we study the statistical mechanics of a system of an
arbitrary number of
entangled polymer rings. Mathematically, two or more entangled
polymers form what is called a
link. Single polymer rings form instead knots.
We will restrict ourselves to systems in the configurations of
$2s-$plats. Roughly speaking, $2s-$plats are knots or links obtained
by braiding together a set of $2s$ strings and connecting their ends
pairwise \cite{birman}. A physical realization of $2s-$plats could be
that of two 
rings topologically entangled together and with some of their points
attached to two membranes or surfaces. In nature $2s-$plats occur for
example in the DNA of 
living organisms \cite{diao,sumners,sumners2,darcy}. Indeed, it is believed
that most knots and links 
formed by DNA are in the class of $4-$plats \cite{sumners}. These
biological 
applications have inspired the research of 
Ref.~\cite{FFPLA2004}, in which $4-$plats have been studied with the
methods of statistical mechanics and field theory. In particular, in
\cite{FFPLA2004} it has been established an
analogy between polymeric $4-$plats and anyons, showing in this way
the tight relations between two component systems of quasiparticles
and the theory of  polymer
knots and links.
After the publication of \cite{FFPLA2004}, 
interesting applications of analogous anyon systems to
topological quantum 
computing have been proposed \cite{dassarma,nayak,wilczek}.
These applications are corroborated by the results of 
experiments concerning  the detection of
anyons obeying a nonabelian statistics, see for example
\cite{goldman}. While these results have appeared in 2005 and are
still under debate
\cite{wilczek,anyonsexp2013}, other systems in which non-abelian
anyon statistics could be present
have been discussed \cite{gurarie,dolev}.

Motivated by these recent advances, we study here the general
case of $2s-$plats in which  $N$ polymer rings are entangled
together
to form a link. 
The topology of the link is distinguished using the Gauss
linking number. This is a weak topological invariant, so that many
inequivalent topological configurations characterized  by some value
of the Gauss linking number are allowed. However, since we are restricting
ourselves to conformations that, by construction, must remain
$2s-$plats, we are implicitly imposing a more stringent topological
condition on the system than that imposed merely by the Gauss linking
number. For example, both the unlink and the Whitehead link have zero
Gauss linking number, but a $4-$plat unlink is not allowed to change
into a Whitehead link, which cannot be realized as a $4-$plat.
Viceversa, a $6-$plat Whitehead link will not transform into a
$4-$plat unlink, despite the fact that both topological configurations
share the same value of the Gauss linking number.

Among all knot and link
configurations, the class of
$2s-$plats is very special. For instance, it is possible to decompose the
trajectory of a $2s-$plat into a set of $2s$ open subtrajectories that
can be further
interpreted as the trajectories of $2s$ polymer chains directed along
an arbitrary direction. 
Without losing generality, we assume that this direction
coincides with the $z-$axis.
Successively, we map into a field theory
 the system of $2s-$directed
polymers resulting from the decomposition of the plat.  After the passage to second
quantized fields,
a model describing a gas of quasiparticles is obtained. In this model, the $z$ coordinate
becomes the
"time", while the monomer densities of the $2s$
directed polymers may be interpreted as  quasiparticle densities of a multi-layered
anyon gas. 
All the nonlocalities and strong nonlinearities of the original theory due to
the topological constraints disappear in the field theoretical formulation.
A remarkable feature of polymers in the configuration of a $2s-$plat
is that these systems admit self-dual points and
their Hamiltonian can be minimized by self-dual
solutions of the classical equations of motion. 
Here we show that in the case of a $4-$plat these solutions may be
explicitly constructed after solving a cosh-Gordon equation.
The
self-dual conformations of a $4-$plat should
be particularly stable and, on the other side,
with the present technologies \cite{membranes} it is possible to realize
polymer $2s-$plats in the laboratory. 
Thus there is some hope that some effect  related  to these
conformations could be
observable.

Apart from the existence of self-dual solutions, the field theoretical
model developed in this work  has also phenomenological consequences
that are
relevant for the statistical mechanics of polymers.
First of all, it provides an explicit and  nonperturbative expression of the
 interactions among  the monomers arising due to the constraints which fix
the topological configuration of the $2s-$plat. After an exact
summation over the abelian BF-fields, it turns out that the monomers
are subjected to forces of topological origin that have a two-body and
a three-body components. 
The two-body interactions can
be both attractive and repulsive, depending on the conformation of the
system and strongly interphere with the non-topological interactions,
which two-body interactions as well.  This results confirms at a
nonperturbative level the outcome of a previous calculation performed with the help of the
method of the effective potential \cite{ferrari}, where it was found
that the monomers of two polymer rings attract themselves due to the
 topological constraints
counterfeiting the excluded volume interactions typical of
polymers in a good solution.
In the particular case of a $4s-$plat, it has been shown in
\cite{FFPLA2004} that,
after a
Bogomol'nyi transformation, it is possible to single out contributions
of the two-body forces of topological
origin that match exactly, apart from
proportionality constants, the excluded volume forces.
What is somewhat  unespected
is the presence of
three-body interactions in a polymer system subjected to topological
constraints imposed with the help of the Gauss linking number. This is
surprising because the Gauss linking number is able to take
into account only the topological relations between pairs of
trajectories. For this reason, one could  expect that this type
of constraints is rather associated with interactions between pairs of
monomers belonging to two different chains. Indeed, before the second
quantization procedure, the explicit expression of the Gauss linking
number can be interpreted as a (nonlocal) two-body potential related
to forces acting on
the bonds located on two different polymers.
Three-body forces give a vanishing contribution in the case of links
with two polymers only, see Ref.~\cite{ferrari}. However, we show here
that there are processes in which three-body forces are relevant if
the number of loops involved in the link is equal to three or higher.

This paper is organized as follows.
Before  mapping the partition function of a general $2s-$plat
into that  of anyons, it is necessary to split the trajectories of the
$N$ polymer rings forming the plat into a set of $2s$
subtrajectories. 
The splitting procedure and the definition of a suitable "time" variable
that parametrizes the $2s$ subtrajectories
is carefully described in Section~\ref{sectionII}.
In Section~\ref{sectionIII} it is shown how it is possible to
implement and simplify in the partition function of the $2s-$plat
 the constraints that fix the possible topological
configurations in which the system  of polymer
rings linked together can be found. The constraints are imposed using the Gauss
linking number. The treatment follows
the method already established in Ref.~\cite{FELAPLB1998}, but its
generalization to
the case in which the trajectories
are splitted into subtrajectories 
parametrized by the special "time" coordinate instead of the usual
arc-lengths is new.
To eliminate the nonlinearities and nonlocalities introduced by the
topological constraints, which necessarily have memory since they must
remember the global geometry of the ring in space, we use a set
of abelian BF-fields. Roughly speaking, these fields generate electromagnetic
type interactions with the monomers and create in this way the necessary
"reaction" forces that forbid the system to escape the constraints.
The BF-field theory is quantized in the non-covariant Coulomb gauge,
because this leads to several simplifications and is very convenient
in order to establish the analogy with anyon systems.
How the "covariance" of the theory is recovered
is shown in Appendix~\ref{appendixB} in the particular case of a
$4-$plat. This example is very helpful to interprete
the meaning of the Gauss linking number in the Coulomb gauge, which is
apparently more related to the winding number of open trajectories than to the Gauss
linking number.
In
Section~\ref{sectionIV}
the passage from first quantized polymer trajectories to second
quantized fields is performed. The case of general interactions
between the monomers is considered. After the second quantization
procedure and the introduction of replica complex scalar fields, the
densities of monomers of the original
polymer rings can be regarded as the densities of 
a system of multilayered gas of quasiparticles.
The topological BF-fields are eliminated by integrating them out from the
partition function.
In Section~\ref{sectionVI} some phenomenological consequences 
on the statistical mechanics of the $2s-$plat
coming from the
field theoretical model obtained in Section~\ref{sectionIV} are
presented. 
In Section~\ref{sectionVII} we limit ourselves to $4-$plats,
switching off the non-topological interactions.
In this particular case, studied in Ref.~\cite{FFPLA2004},
it is known that the Hamiltonian of the $4-$plat is minimized by
self-dual solutions.
Here the classical equations of motion are reduced to
a cosh-Gordon equation. It is shown how the explicit expression of the
classical configurations minimizing the Hamiltonian
of the $4-$plat can be constructed out of the solution of this
cosh-Gordon equation. 
Finally, our conclusions are drawn in
Section~\ref{sectionVIII}. 
\section{Polymers as $2s-$plats}\label{sectionII}
Let's consider $N$ closed loops $\Gamma_1,\ldots,\Gamma_N$ of lengths
$L_1,\ldots,L_N$ respectively in a three-dimensional space with
coordinates $(\boldsymbol r,z)$. 
The vector $\boldsymbol r=(x,y)$ spans the two
dimensional space $\mathbb {R}^2$.
$z$ will play  later on
the role of  time. The $N$ loops will be labeled
using as indices the first letters of the latin alphabet:
$a,b,c,\ldots=1,\ldots,N$.
We will assume  that  $\Gamma_1,\ldots,\Gamma_N$ form a $2s-$plat. For
convenience, we briefly review what is a $2s-$plat. First of all, we
recall that a closed trajectory in the three-dimensional space is
from the mathematical point 
of view a knot, while a system of knots linked together forms
a link. 
\begin{figure}
\centering
\begin{minipage}{7cm}
\includegraphics[width=7.2cm]{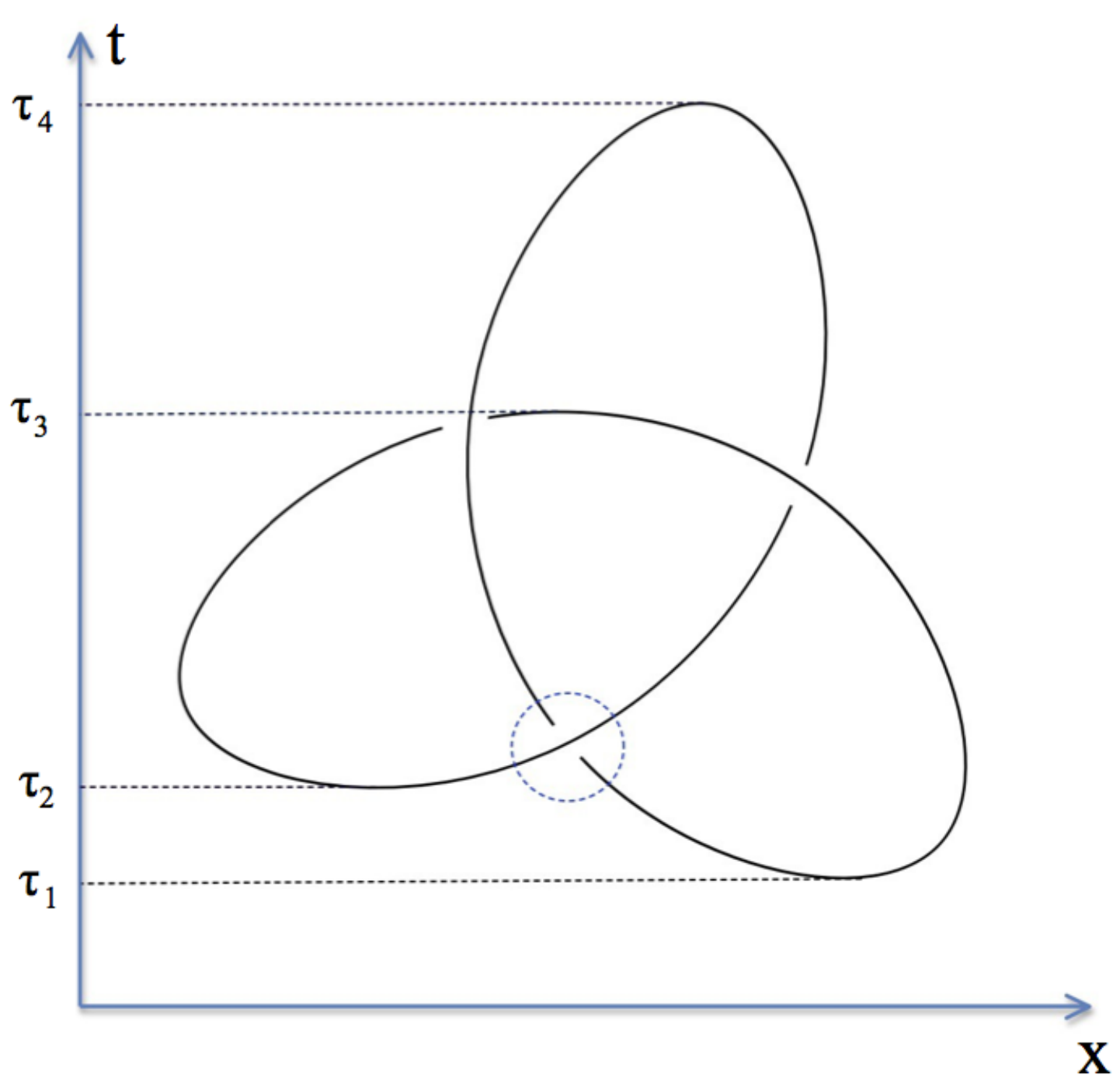} 
\caption{Representation of a trefoil knot in terms as a
  two-dimensional
diagram.}\label{treefoil}
\end{minipage}\hspace{1cm}
\begin{minipage}{6cm}
\vspace*{2.6cm}
\includegraphics[width=5.9cm]{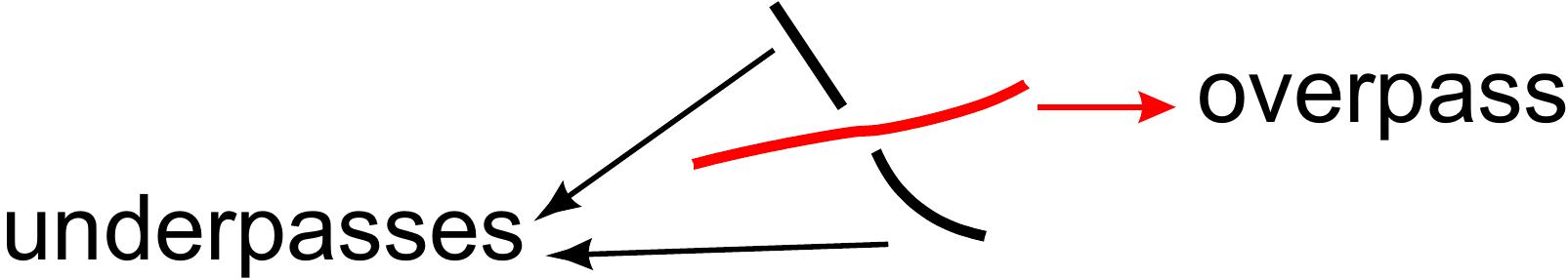}
\vspace{2cm}
\caption{The figure shows one of the crossings which are present
in the diagram of the trefoil knot of Fig.~\ref{treefoil}.}\label{overunder}
\end{minipage}
\end{figure}
After a projection onto a
plane  knots and links may be represented by diagrams like those of
Fig.~\ref{treefoil} and \ref{basiclink}, 
in which the 
original three-dimensional structure is 
simulated by a system of crossings, see Fig.~\ref{overunder}. 
\begin{figure}[bhpt]
\centering
\includegraphics[scale=.25]{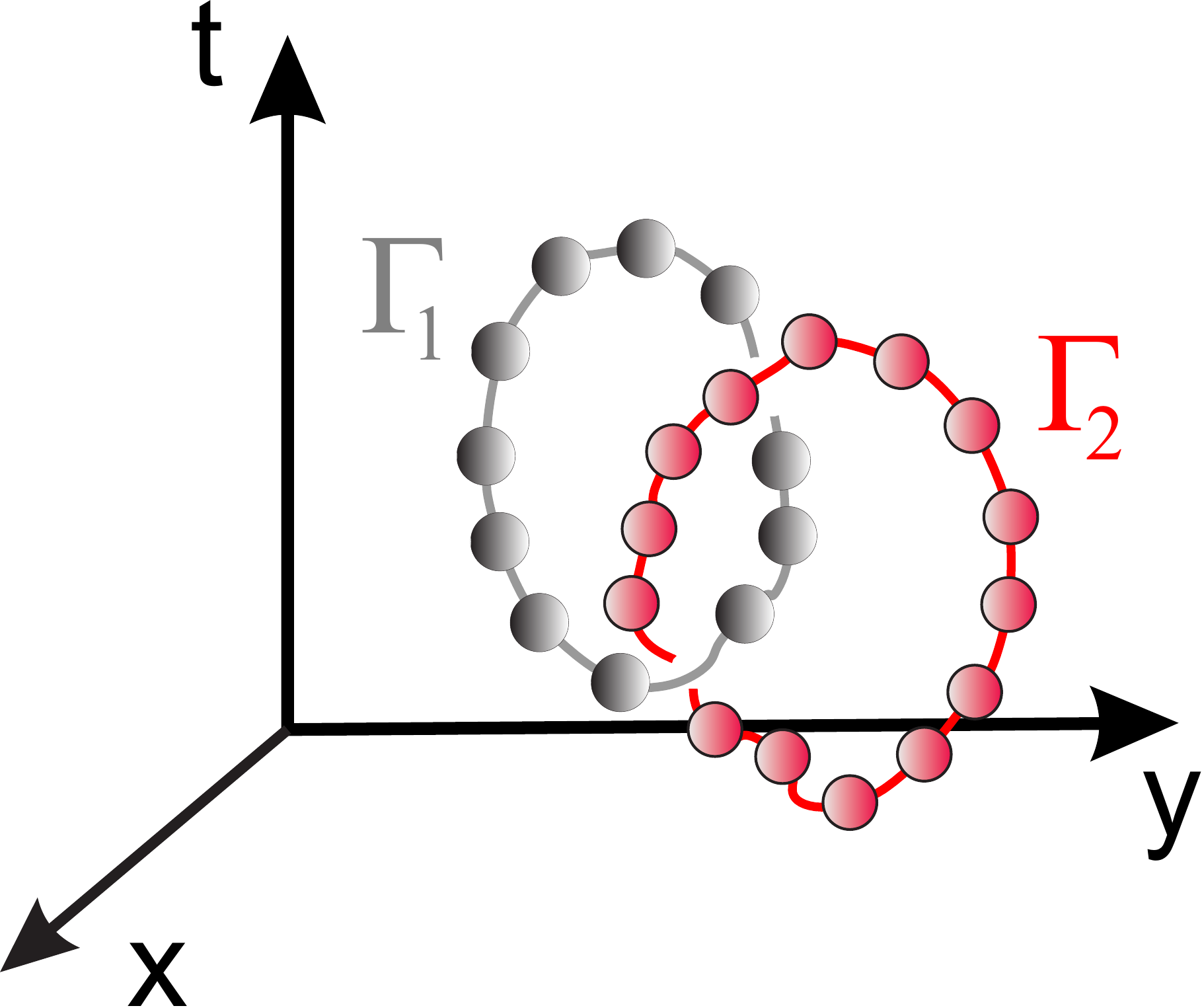}\\
\caption{A link formed by two polymers $\Gamma_1$ and
  $\Gamma_2$.}\label{basiclink} 
\end{figure}
Each crossing is composed by
three arcs, one 
overpass and two underpasses. Giving an orientation to the
trajectories, we can distinguish positive and negative crossings, see
Fig.~\ref{pncrossings}.
\begin{figure}[bhpt]
\centering
\includegraphics[scale=.33]{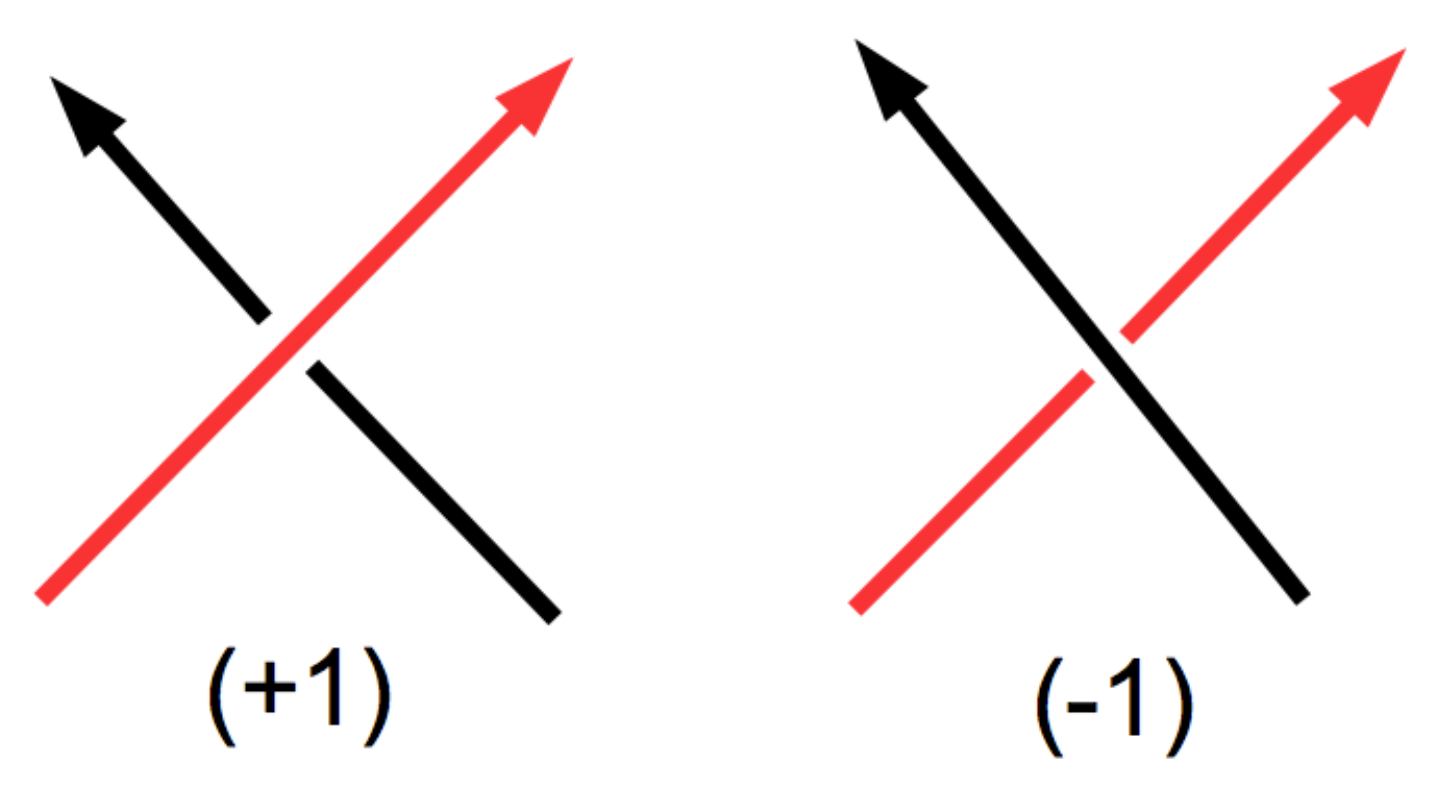}\\
\caption{Left and right crossings of the elements of two oriented
  paths.}\label{pncrossings}  
\end{figure}
One may also realize that the trefoil diagram in Fig.~\ref{treefoil} is
characterized by two minima  and two
maxima. Two dimensional diagrams of this kind, deformed in such a way
that the number $2s$ of minima and maxima is the smallest possible and
the maxima and minima are aligned at the same heights $z_{\max}$
and $z_{\min}$ respectively, are called in knot theory
$2s-$plats~\footnote{Actually, to be rigorous one 
  should still require that 
  neither maxima nor minima occur at the crossing points.}.
The height of a $2s-$plat is measured here with respect
to the $z$ axis.
 In the present case, with some abuse of language,
we will call $2s-$plats 
any system of $N$ three-dimensional knots
realized in such a way that the trajectories of the knots are
characterized by a number $2s$ of maxima and minima. The locations of
the $s$ points of maxima and those of the $s$ points of minima are
fixed, i. e. they are not allowed to fluctuate and their number $2s$
is constant. The points of maxima and minima do not need to be aligned
as it happens in the mathematical definition of a $2s-$plat.
An example of the two-dimensional diagram of such a physical
$2s-$plat is given in Fig.~\ref{basiclink}. 
Let us denote
with the symbols $ \tau_{a,I_a}$, $I_a=0,\ldots, 2s_a-1$,
the heights of the maxima and minima of each
trajectory $\Gamma_a$, for $a=1,\ldots,N$. Of course, it should be
that
\begin{equation}
\sum_{a=1}^Ns_a=s.
\label{conscond}
\end{equation}
We choose $\tau_{a,0}$ to be the height 
of the absolute minimum of each trajectory $\Gamma_a$.
Starting from $\tau_{a,0}$, we select the orientation of
$\Gamma_a$ in such a way that, proceeding along the trajectory
according to that orientation, we will encounter in the order the
points $\tau_{a,1},\tau_{a,2},\ldots,\tau_{a,2s_a}$. Clearly,
$\tau_{a,1}$ is the height of a point of maximum, $\tau_{a,2}$ the height of a
minimum and so on.
Moreover, we should put for consistency
\begin{equation}
\tau_{a,2s_a}\equiv\tau_{a,0}.
\end{equation}
The introduction of two symbols for the same height 
$\tau_{a,0}$ will be useful in the future in order to write
formulas in a more compact form.
In the following, the trajectories $\Gamma_1,\ldots,\Gamma_N$ will be
decomposed into
a set
of directed trajectories
$\Gamma_{a,I_a}$, $a=1,\ldots,N$ and $I_a=1,\ldots, 2s_a$,
whose ends are made to coincide in such a
way that they form the topological configuration of two linked rings.
An example when $s=3$ and $N=1$ is presented in Fig.~\ref{sectioning}.
\begin{figure}[bpht]
\centering
\includegraphics[scale=.5]{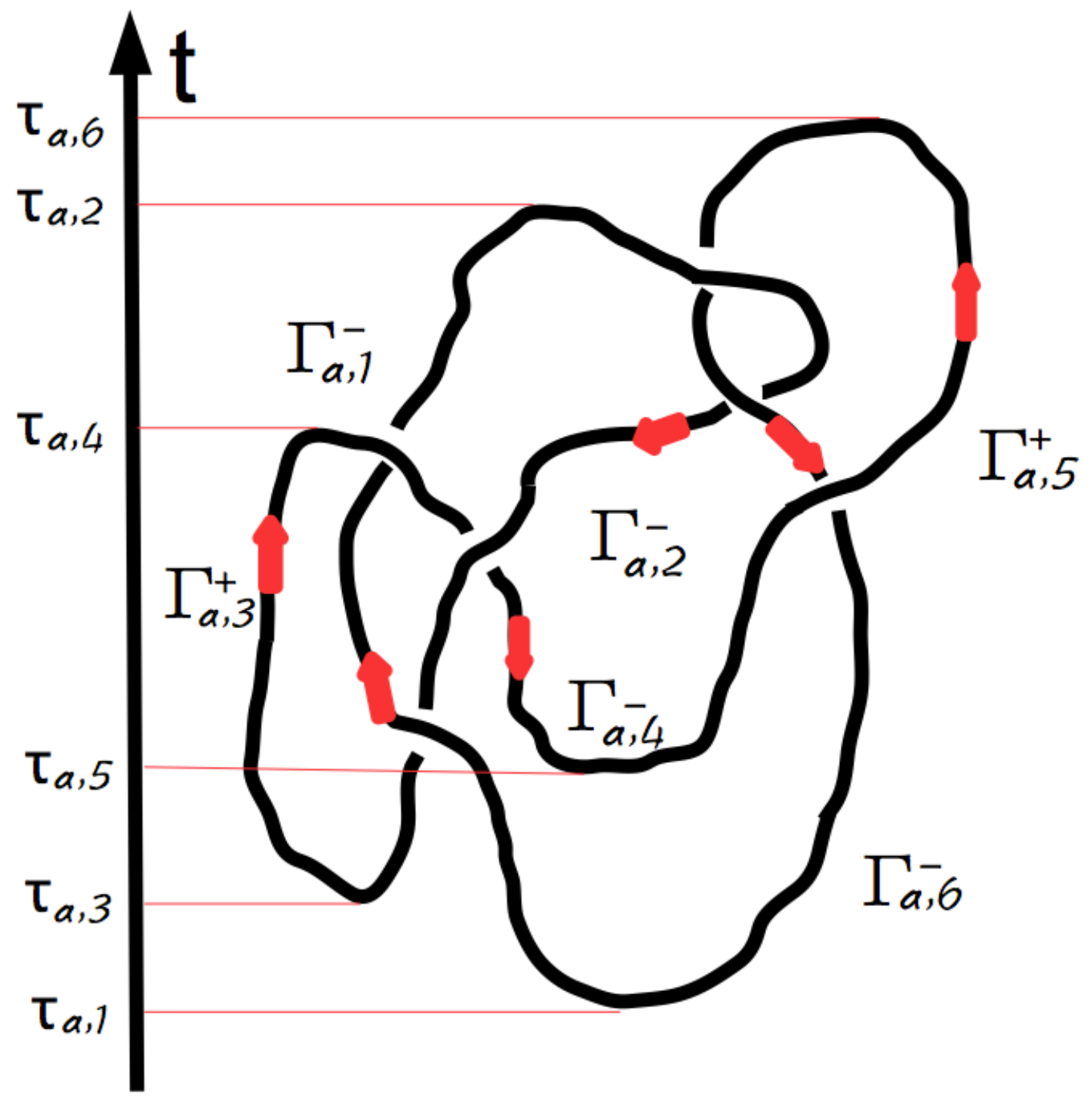}\\
\caption{Sectioning procedure for a $2s$-plat $\Gamma_a$ with
$s=3$.}\label{sectioning}
\end{figure}
In the general case, the set of points belonging to $\Gamma_{a,I_a}$
can be described by the formula:
\begin{equation}
\Gamma_{a,I_a}=\left\{
\boldsymbol r_{a,I_a}(z_{a,I_a})\left|\begin{array}{c}
a=1,\ldots,N;\qquad I_a=1,\ldots,2s_a\\
\left\{\begin{array}{rcl}
\tau_{a,I_a-1}\le z_{a,I_a}\le\tau_{a,I_a}\qquad
I_a\mbox{ odd}\\
\tau_{a,I_a}\le z_{a,I_a}\le\tau_{a,I_a-1}\qquad
I_a\mbox{ even}
\end{array}
\right.
\end{array}
\right.\right\}\label{bcondone}
\end{equation}
where the additional conditions:
\begin{eqnarray}
\boldsymbol r_{a,I_a}(\tau_{a,I_a})&=&\boldsymbol
r_{a,I_a+1}(\tau_{a,I_a})\qquad I_a=1,\ldots,2s_a-1\label{bcondtwoa}\\
\boldsymbol r_{a,1}(\tau_{a,0})&=&\boldsymbol r_{a,2s_a}(\tau_{a,0})
\label{bcondtwo}
\end{eqnarray}
are understood. These conditions are needed in order to
 connect together the subtrajectories $\Gamma_{a,I_a}$ so that 
the loop $\Gamma_a$ is reconstructed.
In Eq.~(\ref{bcondone}) the two-dimensional vector $\boldsymbol
r_{a,I_a}(z_{a,I_a})$ represents the 
projection of the trajectory $\Gamma_{a,I_a}$ onto the plane $x,y$
perpendicular to the $z-$axis.
Let us note that we are using the same indexes $I_a$ to label the
trajectories $\Gamma_{a,I_a}$ and the points $
\tau_{a,I_a}$. However, in the first case $I_a=1,\ldots,2s_a$, while
in the second case we have chosen $I_a=0,\ldots,2s_a-1$.
The range of the indices $I_a$ in the variables $z_{a,I_a}$'s and of
the $t_{a,I_a}$'s that will be defined later
is the same as that of the indices labeling the trajectories
$\Gamma_{a,I_a}$'s, i.~e. $I_a=1,\ldots,2s_a$.

We notice that the  $z_{a,I_a}$'s are always
growing. In this way, the fact that
the whole chain is continuous and has a given orientation is not taken
into account. 
Better variables,  respecting both the continuity and orientation of
the trajectories $\Gamma_{a,I_a}$, are 
 the  $t_{a,I_a}$'s, which are defined as follows:
\begin{eqnarray}
t_{a,I_a}=z_{a,I_a}&\qquad&\mbox{when }I_a\mbox{ is
  odd}\label{transfone}\\ 
t_{a,I_a}=-(z_{a,I_a}-\tau_{a,I_a})
+ \tau_{a,I_a-1}
&\qquad&\mbox{when }I_a\mbox{ is even}.\label{transftwo}
\end{eqnarray}
Assuming for instance that $I_a$ is odd, then for any two consecutive
trajectories $\Gamma_{a,I_a}$ and $\Gamma_{a,I_{a+1}}$ the range of
the variables $t_{a,I_a}$ and $t_{a,I_{a+1}}$ is given by:
\begin{equation}
\tau_{a,I_a-1}\le t_{a,I_a}\le\tau_{a,I_a}\qquad I_a\mbox{ odd}\qquad
1\le I_a\le 2s_a-1
\end{equation}
Instead, if $I_a$ is even:
\begin{equation}
\tau_{a,I_a-1}\ge t_{a,I_a}\ge\tau_{a,I_a}\qquad I_a\mbox{ even}\qquad
2\le I_a\le 2s_a.
\end{equation}
Let us recall that by our conventions the trajectories labeled by odd
$I_a$'s are oriented from a point of minimum to a point of maximum,
while trajectories with even values of $I_a$ go from a point of
maximum to a point of minimum. Accordingly, the new variables
$t_{a,I_a}$ have been 
chosen in such a way that they increase from the minimum to the
maximum when $I_a$ is odd, while they decrease from the point of
maximum to that of minimum when $I_a$ is even.
Finally, we provide the definition of the curves $\Gamma_{a,I_a}$
parametrized with the help of the $t_{a,I_a}$'s:
\begin{equation}
\Gamma_{a,I_a}=\left\{
\boldsymbol r_{a,I_a}( t_{a,I_a})\left|\begin{array}{c}
a=1,\ldots,N;\qquad I_a=1,\ldots,2s_a\\
\left\{\begin{array}{rcl}
\tau_{a,I_a-1}\le  t_{a,I_a}\le\tau_{a,I_a}\qquad
I_a\mbox{ odd}\\
\tau_{a,I_a-1}\ge  t_{a,I_a}\ge\tau_{a,I_a}\qquad
I_a\mbox{ even}
\end{array}
\right.
\end{array}
\right.\right\}\label{conddd}
\end{equation}
Of course, the boundary conditions \ceq{bcondtwoa} and \ceq{bcondtwo} are
always understood.

The variables $t_{a,I_a}$ arise in a natural way when a curvilinear
integral around the loop $\Gamma_a$ is split into many
subtrajectories $\Gamma_{a,I_a}$. In fact, let's consider for example
integrals of the kind
\begin{equation}
I=\oint_{\Gamma_a} d{\tilde x}^\mu_a(d_a)A_\mu({\tilde
  x}_a(d_a))
\label{integrali}
\end{equation}
where
 the symbol
${\tilde x}^\mu_a(d_a)=({\tilde{\boldsymbol r}_a}(d_a),{\tilde
  x}_a^3(d_a))$ 
 denotes the points of the trajectory $\Gamma_a$
parametrized in terms of the arc-length $d_a$, $0\le d_a\le
L_a$.
$A_\mu({\tilde x}_a(d_a))$ is an abelian gauge field on ${\mathbb
  R}^3$.
It is easy to show that, after splitting the loop $\Gamma_a$ into the
subtrajectories $\Gamma_{a,I_a}$, on each of these subtrajectories it
is possible to change the arc-length $d_a$ with the parameters
$t_{a,I_a}$. If one does that, the curvilinear integral $I$ of
Eq.~(\ref{integrali}) becomes parametrized by the variables
$t_{a,I_a}$ and may be expressed as follows
\begin{equation}
I=\sum_{I_a=1}^{2s_a}\int_{\tau_{a,I_a-1}}^{\tau_{a,I_a}}\left[
\frac{d\boldsymbol r_{a,I_a}(t_{a,I_a})}{dt_{a,I_a}}\cdot \boldsymbol
A(\boldsymbol r_{a,I_a}(t_{a,I_a}),t_{a,I_a}) + A_3(\boldsymbol
r_{a,I_a}(t_{a,I_a}),t_{a,I_a})
\right]\label{integralinewpar}
\end{equation}
where
\begin{equation}
t_{a,I_a}={\tilde x}^3_a(d_a)\qquad\qquad \boldsymbol
r_{a,I_a}(t_{a,I_a})=\boldsymbol r_{a,I_a}({\tilde x}_a^3(d_a))=
{\tilde{\boldsymbol r}_a}(d_a).\label{varexch}
\end{equation}
Of course, the above equation is valid only if $d_a$ is restricted
on the trajectory $\Gamma_{a,I_a}$, i.e., $\delta_{a,I_a-1}\le
d_a\le\delta_{a,I_a}$. The $\delta_{a,I_a}$'s denote the values of the
arc-length at
the points of maxima and minima of the $2s_a-$plat
$\Gamma_a$. Clearly, ${\tilde x}_a^3(\delta_{a,I_a})=\tau_{a,I_a}$.
\section{Fixing the topological properties of a $2s-$plat:
the case of the Gauss linking number}\label{sectionIII}
In the case of a $2s-$plat composed by $N$ loops
$\Gamma_1,\ldots,\Gamma_N$, it is possible to specify the winding
number between any two subtrajectories $\Gamma_{a,I_a}$ and
$\Gamma_{b,I_b}$ composing the plat. These winding numbers cannot
change due to the thermal fluctuations, because the end points
$(\boldsymbol r(\tau_{a,I_a-1}), \tau_{a,I_a-1})$ 
and $(\boldsymbol r(\tau_{a,I_a}), \tau_{a,I_a})$ 
 of each
subtrajectory $\Gamma_{a,I_a}$ must be fixed in our construction. This
fact can be used to constrain the $2s-$plat to 
stay in very complex topological configurations. In the following,
however, we will not adopt this strategy. The topological
configurations of the system will rather be imposed 
as in Refs.~\cite{FELAPLB1998} by
applying the Gauss linking number. 
\subsection{The standard approach of imposing the constraints with the
Gauss linking number}
The Gauss linking number is a link invariant expressing the
topological states of two closed trajectories linked together. Due to
the fact that it can only be applied to pairs of loops, here we
restrict ourselves
 for simplicity to the case of a $2s-$plat composed by only two loops
 $\Gamma_1$ and $\Gamma_2$. Note that  each of these two loops is
a plat too having $s_a$ points of maxima and $s_a$ points of
minima with $a=1,2$. For consistency, it should be that $s=s_1+s_2$. 
The  Gaussian linking
number is defined as follows
\begin{equation}
\chi(\Gamma_{1},\Gamma_{2})=\frac{1}{4\pi}\epsilon_{\mu\nu\rho}
\oint_{\Gamma_{1}} d\tilde x^{\mu}_{1}(d_1)\oint_{\Gamma_{2}}
 d\tilde x^{\nu}_{2}(d_2)\frac{(\tilde x_{1}(d_1)
-\tilde x_{2}(d_2))^{\rho} 
}{|\tilde x_{1}(d_1)-\tilde x_{2}(d_2)|^{3}} \label{ln}
\end{equation}
where the $\tilde x^\mu_a(d_a)$'s and the arc-lengths $d_a$'s, $a=1,2$
have been already 
defined at the end of the previous Section, after
Eq.~(\ref{integrali}). 
The trajectories of the two loops will be
topologically constrained
by the condition
\begin{equation}
m_{12}=\chi(\Gamma_1,\Gamma_2)\label{topconst}
\end{equation}
$m_{12}$ being a given integer. The above constraint is imposed by
inserting
the Dirac delta function
$\delta(m_{12}-\chi(\Gamma_1,\Gamma_2))$ in
the partition function of the $2s-$plat, where
the statistical sum over all conformations of $\Gamma_1$ and
$\Gamma_2$ is performed. Of course, the analytical treatment of such a
delta function in a path integral is difficult. Some
simplification is obtained by 
passing to
the Fourier representation
\begin{equation}
\delta(m_{12}-\chi(\Gamma_1,\Gamma_2))=\int_{-\infty}^{+\infty}
\frac{d\lambda_{12}}{\sqrt{2\pi}} 
\;{\rm e}^{-i\lambda_{12}(m_{12}-\chi(\Gamma_1,\Gamma_2))}.\label{deltaft}
\end{equation}
Even in the Fourier representation,
the difficulty of having to deal with the Gauss linking number in the
exponent appearing in the right hand side of Eq.~(\ref{deltaft})
remains. Formally,
this link invariant introduces a term that  resembles
the potential of a two-body interaction which is
 both nonlocal and 
nonpolynomial. For this reason, the treatment of the Gauss
linking number in any microscopical model of topologically entangled
polymers is complicated. The best strategy to deal with this problem
 consists
in rewriting the delta
function
$\delta(m_{12}-\chi(\Gamma_1,\Gamma_2))$ as a correlation function of the
holonomies of a local field theory, namely the so-called abelian
BF-model \cite{FELAPLB1998,FELAJPA1999,thompsonblau}
\begin{equation}
\delta(m_{12}-\chi(\Gamma_1,\Gamma_2))=\int_{-\infty}^{+\infty}d\lambda_{12}
\;{\rm e}^{-i\lambda_{12} m_{12}}{{\cal Z}}_{\rm BF}(\lambda_{12})\label{maintop}
\end{equation}
where
\begin{eqnarray}
{{\cal Z}_{\rm BF}}(\lambda_{12})&=&\int{\cal D} B_\mu^{12}(x){\cal
  D} C_\mu^{12}(x)
\;{\rm e}^{-iS_{\rm BF}\left[B,C\right]}\nonumber\\
&\times&
\;{\rm e}^{-i\tilde c_{12}\oint_{\Gamma_1}d\tilde x^\mu_1(d_1)
  B_\mu^{12}(\tilde{x}_1(d_1))}
\;{\rm e}^{
-i\tilde d
\oint_{\Gamma_2}
d\tilde x^\mu_2
(d_2)C_\mu^{12}(\tilde{x}_2(d_2))
}.\label{c2a3}
\end{eqnarray}
In the above equation we have put $x\equiv(\boldsymbol x,t)$ to be
dummy integration variables spanning the whole three-dimensional space
$\mathbb R^3$. Moreover,
$S_{\rm BF}[ B, C]$ denotes the action of the abelian BF-model
\begin{equation}
S_{\rm BF}[B, C]=\frac\kappa{4\pi}\int d^3x
B_\mu^{12}(x)\partial_\nu  C_\rho^{12}(x)\epsilon^{\mu\nu\rho}. \label{bfmodelaction}
\end{equation}
Above $\epsilon^{\mu\nu\rho} $, $\mu,\nu,\rho=1,2,3$, is the completely
antisymmetric 
$\epsilon-$tensor density defined by the condition $\epsilon^{123}=1$.
$\kappa$ is the coupling constant of the BF-model. 
Finally, 
the constants $\tilde c_{12}$ and
$\tilde d$ are given by:
\begin{equation}
\tilde c_{12}=\lambda_{12}\qquad\qquad \tilde d=\frac\kappa{8\pi^2}.
\end{equation}
While there is some freedom in choosing $\tilde c_{12}$ and $\tilde d$,
one unavoidable
requirement in order that Eq.~(\ref{maintop}) will be satisfied is that one of
these parameters should be linearly dependent on $\kappa$.
In this way, it is easy to check that $\kappa$ may be completely
eliminated from Eq.~(\ref{c2a3}) by performing a
rescaling of one of
the two fields $B_\mu^{12}$ and $ C_\mu^{12}$.
This is an expected result, because $\kappa$
does not appear in
the left hand 
side of Eq.~(\ref{maintop}), so that it
cannot be a new parameter of the theory.
By introducing the currents:
\begin{equation}
\zeta_{12}^\mu( x)=\tilde
c_{12}\oint_{\Gamma_1}d\tilde x_1^\mu (d_1)\delta^{(3)}(
x- \tilde{ x}_1(d_1))
\qquad\qquad
\xi_{12}^\mu( x)=\tilde
d\oint_{\Gamma_2}d\tilde x_2^\mu (d_2)\delta^{(3)}(
x- \tilde{ x}_2(d_2))
\label{tildecurrent}
\end{equation}
$ {\cal Z}_{\rm BF}(\lambda_{12})$ may be rewritten in the more compact way:
\begin{eqnarray}
{\cal Z}_{\rm BF}(\lambda_{12})&=& \int{\cal D} B_\mu^{12}(x)
{\cal D} C_\mu^{12}(x)\;{\rm e}^{-iS_{\rm BF}[ B, C]}
\;{\rm e}^{-i\int d^3x\left[
\zeta^\mu_{12}
(x) B_\mu^{12}(x)+
\xi_{12}^\mu
(x) C_\mu^{12}(x)
\right]}.\label{maintopcurrents}
\end{eqnarray}
With Eq.~(\ref{maintopcurrents}) the goal of transforming the
nonlinear and nonlocal interaction appearing in the right hand side of
Eq.~(\ref{deltaft}) is achieved. The right hand side of
Eq.~(\ref{maintopcurrents}) represents in fact a local field theory,
the BF-model, interacting with the trajectories $\Gamma_1$ and $\Gamma_2$.
Of course, the price paid for that simplification is the introduction
of the fields 
$B_\mu^{12}$ and $C_\mu^{12}$.
\subsection{How to impose constraints on a link composed by plats
  using the Gauss linking number}
In all the above discussion, the two trajectories
$\Gamma_1$ and $\Gamma_2$ have been
parametrized with the help of the arc-lengths $d_1$ and $d_2$.
However,
in the present case the loops $\Gamma_1,\ldots,\Gamma_N$ are realized
as a set of open 
paths $\Gamma_{a,I_a}$ connected together by the conditions
(\ref{bcondtwoa})--(\ref{bcondtwo}). 
The subtrajectories $\Gamma_{a,I_a}$'s are
directed paths
$\boldsymbol
r_{a,I_a}(t_{a,I_a})=(x^1_{a,I_a}(t_{a,I_a}),x^2_{a,I_a}(t_{a,I_a}))$
parametrized
by the variables $t_{a,I_a}$.
This difference of parametrization introduces several important
changes. Apart from the fact that we have to deal with many
subtrajectories, also one degree of freedom, represented by the third
coordinate $x^3_a(s_a)$, disappears due to the change (\ref{varexch}).
As a consequence, the  method illustrated in the previous Subsection
in order  to
express the Gauss linking number as an amplitude of the BF-model, in particular
Eq.~(\ref{maintop}),  should be
changed appropriately. Thus, we rewrite the partition
function ${\cal Z}_{\rm BF}(\lambda_{12})$ of Eq.~(\ref{c2a3}) 
using the variables $t_{a,I_a}$ to parametrize the subtrajectories
$\Gamma_{a,I_a}$. The way in which the curvilinear integrals along the
loops $\Gamma_1$ and $\Gamma_2$ appearing in Eq.~(\ref{c2a3})
should be replaced by integrals 
over the subtrajectories $\Gamma_{a,I_a}$
is shown in Eqs.~(\ref{integrali}) and (\ref{integralinewpar}).
As a result, we arrive at the following expression of the partition
function ${\cal Z}_{\rm BF}(\lambda_{12})$:
\begin{eqnarray}
{\cal Z}_{\rm BF}(\lambda_{12})&=& \int{\cal D} B_\mu^{12}(x)
{\cal D} C_\mu^{12}(x)\;{\rm e}^{-S_{\rm BF}[ B, C]}
 \;{\rm e}^{-i\int d^3x\left[
\boldsymbol\zeta_{12}(\boldsymbol x,t)\cdot \boldsymbol B^{12}(\boldsymbol x,t)+
\zeta^3_{12}(\boldsymbol x,t) B_3^{12}(\boldsymbol x,t)
\right]}\nonumber\\
&\times& \;{\rm e}^{-i\int d^3x\left[
\boldsymbol\xi_{12}(\boldsymbol x,t)\cdot \boldsymbol C^{12}(\boldsymbol x,t)+
\xi^3_{12}(\boldsymbol x,t) C_3^{12}(\boldsymbol x,t)
\right]}
\label{zlambdawotilde}
\end{eqnarray}
where $S_{\rm BF}[ B, C]$
coincides with the action (\ref{bfmodelaction}) and
\begin{eqnarray}
\boldsymbol \zeta_{12}(\boldsymbol x,t)&=&\tilde
c_{12}\sum_{I_1=1}^{2s_1}\int_{\tau_{1,I_1-1}}^{\tau_{1,I_1}}dt_{1,I_1}
\dot{\boldsymbol r}_{1,I_1}(t_{1,I_1})
\delta^{(2)}(\boldsymbol x-\boldsymbol r_{1,I_1}(t_{1,I_1}))
\delta(t-t_{1,I_1})
\label{currentwotilde1}\\
\boldsymbol \xi_{12}(\boldsymbol x,t)&=&\tilde
d\sum_{I_2=1}^{2s_2}\int_{\tau_{2,I_2-1}}^{\tau_{2,I_2}}dt_{2,I_2}
\dot{\boldsymbol r}_{2,I_2}(t_{2,I_2})
\delta^{(2)}(\boldsymbol x-\boldsymbol r_{2,I_2}(t_{2,I_2}))
\delta(t-t_{2,I_2})
\label{currentwotilde2}\\
 \zeta_{12}^3(\boldsymbol x,t)&=&\tilde
c_{12}\sum_{I_1=1}^{2s_1}\int_{\tau_{1,I_1-1}}^{\tau_{1,I_1}}dt_{1,I_1}
\delta^{(2)}(\boldsymbol x-\boldsymbol r_{1,I_1}(t_{1,I_1}))
\delta(t-t_{1,I_1})
\label{currentwotilde3}\\
\xi_{12}^3(\boldsymbol x,t)&=&\tilde
d\sum_{I_2=1}^{2s_2}\int_{\tau_{2,I_2-1}}^{\tau_{2,I_2}}dt_{2,I_2}
\delta^{(2)}(\boldsymbol x-\boldsymbol r_{2,I_2}(t_{2,I_2}))
\delta(t-t_{2,I_2}).
\label{currentwotilde4}
\end{eqnarray}
\subsection{The Coulomb gauge}
Now we use the Fourier representation of the topological constraints of
 Eq.~(\ref{maintop}), but with the partition function
${\cal Z}_{\rm BF}(\lambda_{12})$ written in the form of
Eq.~(\ref{zlambdawotilde}). 
In  this way
the path
integral over all conformations of the $2s-$plat
can be split into  path integrals over all conforations of the
subtrajectories $\Gamma_{a,I_a}$. The latter can be regarded as the
trajectories
 of a two-dimensional system
of $2s$ particles interacting with
abelian BF fields. 
In order to establish an explicit analogy between polymers and
two-dimensional particles evolving in time,  it is convenient
to choose a 
non-covariant gauge like the Coulomb gauge.
Similar approaches like that proposed here can be found in
  \cite{froehlichking,labastida}. Interestingly, in \cite{labastida}
  Chern-Simons 
  field theories quantized in noncovariant gauges have also been applied
  to express the 
  knot and link invariants of $2s-$plats,  called in
  \cite{labastida}  Morse knots.
In Refs.~\cite{froehlichking} and \cite{labastida} 
knots and links are however static, they do not fluctuate, and
the calculations have been performed in noncovariant 
 gauges different from the Coulomb gauge.

To begin with, we impose the Coulomb gauge condition on the $B$ and $C$ fields
\begin{equation}
\partial^{i}B_{i}^{12}=\partial^{i}C^{12}_i=0 \label{GG}
\end{equation}
where $i=1,2$ labels the  first two components  of the vector
potentials $B_\mu^{12}=(\boldsymbol B^{12},B_3^{12})$
and $C_\mu^{12}=(\boldsymbol C^{12},C_3^{12})$.
After the gauge choice (\ref{GG}), the action of the BF model
(\ref{bfmodelaction}) becomes
\begin{equation}
S_{\rm BF,CG}\left[B,C\right]=\frac
{\kappa}{4\pi}\int\! d^{3}x\left[B^{12}_3
\epsilon^{ij}\partial_{i}C^{12}_{j}+C_3^{12}
\epsilon^{ij}\partial_{i}B_{j}^{12}\right] \label{CSCG}
\end{equation}
with $\epsilon^{ij}=\epsilon^{ij3}$ being the two-dimensional
completely antisymmetric tensor.
The gauge fixing term vanishes in the pure Coulomb gauge where the
conditions \ceq{GG} are strictly satisfied. Also the Faddeev-Popov term,
which in principle should be 
present in Eq.~\ceq{CSCG}, may be neglected because the ghosts
decouple from all other fields. 

The requirement of transversality of (\ref{GG}) in the "spatial"
directions $x^1,x^2$
implies that the  components $B_{i}^{12}$ and $C_{i}^{12}$ 
of the BF fields
may be expressed in terms of
two scalar fields $b^{12}$ and $c^{12}$ via the Hodge decomposition:
\begin{equation}
B_{i}^{12}=\epsilon_{ij}\partial^{j}b^{12}\qquad\qquad
C_{i}^{12}=\epsilon_{ij}\partial^{j}c^{12}. \label{hodgedef}
\end{equation}
After performing the above substitutions of fields in the BF action of
Eq.~\ceq{CSCG}, we obtain
\begin{equation}
S_{\rm BF,CG}[B,C]=\frac
{\kappa}{4\pi}\int d^{3}x[B_3^{12}\Delta c^{12}+C_3^{12}\Delta
  b^{12}]. \label{bcaction} 
\end{equation}

Now we compute  the propagator of the BF fields
\begin{equation}
G_{\mu\nu}(\boldsymbol x,t;\boldsymbol
y,t^{\prime})=\langle B_{\mu}^{12}(\boldsymbol x,t),C_{\nu}^{12}(\boldsymbol
y,t^{\prime})\rangle.
\end{equation}
Only the following components of the
propagator are different from zero:
\begin{equation}
G_{3i}(\boldsymbol x,t;\boldsymbol y,t^{\prime})
=\frac{\delta(t-t^{\prime})}{2\kappa}\epsilon_{ij}
\partial_{\boldsymbol y}^{j}\log|
\boldsymbol x-\boldsymbol y|^{2} \label{propone}
\end{equation}
\begin{equation}
G_{i3}(\boldsymbol x,t;\boldsymbol
y,t^{\prime})=-G_{3i}(
\boldsymbol x,t;\boldsymbol y,t^{\prime}).\label{proptwo}
\end{equation}
The path integration
over the scalar
fields 
$b^{12}$ and $c^{12}$
in
the partition function ${\cal Z}_{\rm BF}(\lambda)$ 
is gaussian and can be performed analytically eliminating completely
the gauge fields.
A natural question that arise at this point is the interpretation
of the topological constraint
\ceq{topconst} in the Coulomb gauge. As a matter of fact, the
BF propagator in the Coulomb gauge breaks explicitly the invariance of
the BF model under general three-dimensional transformation.
It seems thus
hard to recover the form (\ref{ln}) of the Gauss
linking number in 
this gauge.
Of course, an
equivalent constraint should be obtained in the Coulomb gauge due to
gauge invariance. 
In Appendix~\ref{appendixB} it will be shown by a direct calculation 
in the case
of a $4-$plat
that this is actually true. The computation of the expression of the
equivalent of the
Gauss linking number in the Coulomb gauge for a general $2s$-plat is
however technically complicated and will not be performed here.

\section{The partition function of a plat}\label{sectionIV}
\subsection{Directed polymers with topological constraints}
In order to write the partition function of a $2s-$plat, we follow the
strategy explained in the previous
Section of dividing each trajectory $\Gamma_a$, $a=1,\ldots,N$, into $2s_a$ open paths
$\Gamma_{a,I_a}$, $I_a=1,\ldots,2s_a$.
The statistical sum ${\cal Z}_{\rm pol}(\{m\})$ of the system is performed
over all
 conformations $\boldsymbol r_{a,I_a}(t_{a,I_a})$
of the subtrajectories $\Gamma_{a,I_a}$ using path integral methods, i.e.: 
\begin{equation}
{\cal Z}_{\rm pol}(\{m\})=
\int\limits_{\mbox{\scriptsize boundary}\atop{\mbox{\scriptsize
 conditions}}} \!\left[\prod_{a=1}^N\prod_{I_a=1}^{2s_a}{\cal D}\boldsymbol
r_{a,I_a}(t_{a,I_a})\right]\,{\rm e}^{-(S_{\rm free}+S_{\rm EV})}
\prod_{a=1}^{N-1}\prod_{b=a+1}^N\delta\left(
m_{ab}-\chi(\Gamma_a,\Gamma_b)
\right).
\label{ppm0}
\end{equation}
In the above equation the boundary conditions on the trajectories 
$\boldsymbol
r_{a,I_a}(t_{a,I_a})$ enforce the 
constraints
\ceq{bcondtwoa} and \ceq{bcondtwo}. 
The free part of the action $S_{\rm free}$ is given by
\begin{equation}
S_{\rm free}=\sum_{a=1}^{N}
\sum_{I_a=1}^{2s_{a}}\int_{\tau_{a,I_a-1}}^{\tau_{a,I_a}}\!dt_{a,I_a}(-1)^{I_a-1} 
g_{a,I_a} 
\left|\frac{d\boldsymbol r_{a,I_a}(t_{a,I_a})}{dt_{a,I_a}}\right|^{2}.\label{sfree}
\end{equation}
The parameters $g_{a,I_a}>0$ are
proportional to the inverse of the Kuhn lengths of the trajectories
$\Gamma_{a,I_a}$. 
They are also related 
to the total lengths of the trajectories $\Gamma_{a,I_a}$ according to
the formula provided in
Appendix~\ref{appendixA}.
Let us note that 
$S_{\rm free}$ is a positive definite
functional thanks to the factors $(-1)^{I_a-1}$, which compensate the
fact that the increment $dt_{a,I_a}$ is negative when $I_a$ is even.
The contribution $S_{\rm EV}$ to the total action 
 takes into account the
interactions between the monomers which arise because we treat the
subtrajectories $\Gamma_{a,I_a}$ as directed paths moving in a random
media.
The mechanism through which these interactions appear after the
integration over the non-white random noises is explained in
Ref.~\cite{Kardar}. 
Explicitly, $S_{\rm EV}$ is given by
\begin{eqnarray}
S_{\rm EV}&=&\frac
12\sum_{a=1}^N\sum_{b=1}^N\sum_{I_a=1}^{2s_a}\sum_{I_b=1}^{2s_b}\int_{\tau_{a,I_a-1}}^{\tau_{a,I_a}}
dt_{a,I_a}\int_{\tau_{b,I_b-1}}^{\tau_{b,I_b}}dt_{b,I_b}(-1)^{I_a+I_b-2}
{\cal M}^{a,I_a;b,I_b}
\nonumber\\
&&V\left(\boldsymbol
r_{a,I_a}(t_{a,I_a})-
\boldsymbol
r_{b,I_b}(t_{b,I_b})\right)\delta\left(t_{a,I_a}-t_{b,I_b}\right)
\label{SEV}
\end{eqnarray}
where
\begin{eqnarray}
{\cal M}^{a,I_a;b,I_b}=
\left\{
\begin{array}{l}
0 \qquad\mbox{if $a=b$ and $I_a=I_b$}\\
1 \qquad\mbox{otherwise}
\end{array}
\right. \label{matrixm}
\end{eqnarray}
Due to the matrix ${\cal M}^{a,I_a;b,I_b}$ the interactions between a
subtrajectory with itself are forbidden.
We note that the presence of the delta functions
$\delta\left(t_{a,I_a}-t_{b,I_b}\right)$ 
is necessary to express the fact that the trajectories $\Gamma_{a,I_a}$
and $\Gamma_{b,I_b}$ for $I_a\ne I_b$ may interact only if both $t_{a,I_a}$ 
and $t_{b,I_b}$ belong to the common interval 
$[\tau_{a,I_a-1},\tau_{a,I_a}]\cap [\tau_{b,I_b-1},\tau_{b,I_b}]$.
The potential $V(\boldsymbol
r)$ can be any two-body potential. 
If the random noises are gaussianly distributed as in Ref.~\cite{Kardar},
then
\begin{equation}
V(\boldsymbol r)= V_{0}\delta(\boldsymbol r) \label{twobody}
\end{equation}
$V_{0}$ being a positive constant. 
Again, the factors $(-1)^{I_a+I_b-2}$
appearing in $S_{\rm EV}$ are necessary in order to compensate the fact that
the increments $dt_{a,I_a}$ and $dt_{b,I_b}$ are negative for even values 
of $I_a$ and $I_b$ respectively.
Finally,
the Dirac delta functions inserted in the right hand side of Eq.~(\ref{ppm0})
impose the topological constraints on each pair of trajectories $(\Gamma_a, \Gamma_b)$, $a=1,\ldots,N-1$, $b=a+1,\ldots,N$. 

\subsection{Passage to Field Theory I: the topological states}
According to Eq.~(\ref{maintop}), the physically relevant contributions 
coming from the topological conditions $m_{ab}=\chi(\Gamma_a,\Gamma_b)$ ,
$a=1,\ldots,N-1$, $b=a+1,\ldots,N$, are encoded in the Fourier transform  
 ${\cal Z}_{\rm pol}(\{\lambda\})$ of
the original probability function ${\cal Z}_{\rm pol}(\{m\})$.  Notice that ${\cal
  Z}_{\rm pol}(\{\lambda\})$ is obtained from  ${\cal Z}_{\rm pol}(\{m\})$ 
 by the relation
 \begin{equation}
{\cal Z}_{\rm pol}(\{m\})=\prod_{a=1}^{N-1}\prod_{b=a+1}^{N}
\int_{-\infty}^{+\infty} d\lambda_{ab}\, {\rm e}^{-i\lambda_{ab} m_{ab}} \,{\cal
  Z}_{\rm pol}(\{\lambda\}) .
\label{ggg}
\end{equation}
It is easy to realize that
\begin{eqnarray}
{\cal Z}_{\rm pol}(\{\lambda\})&=&\int\!\left[\prod_{a=1}^{N-1}\prod_{b=a+1}^{N}
{\cal D}B_\mu^{ab}{\cal D}C_\mu^{ab}\right]
{\rm e}^{- i S_{\rm BF}}\nonumber\\
&&\int\limits_{\mbox{\scriptsize boundary}\atop{\mbox{\scriptsize
    conditions}}} \left[\prod_{a=1}^N\prod_{I_a=1}^{2s_a}
\!{\cal D}\boldsymbol r_{a,I_a}(t_{a,I_a})\right]{\rm e}^{-(S_{\rm free}+S_{\rm EV}+S_{\rm top})}
\label{ppm}
\end{eqnarray}
where
\begin{equation}
 S_{\rm BF}=\sum_{a=1}^{N-1}\sum_{b=a+1}^N
\frac{\kappa}{4\pi}\int\!d^{3}xB_{\mu}^{ab}(x)\partial_{\nu}C_{\rho}^{ab}
(x)\epsilon^{\mu\nu\rho}
\label{CSaction}
\end{equation} 
and
\begin{eqnarray}
 S_{\rm top}=i\sum_{a=1}^{N-1}\sum_{b=a+1}^N\lambda_{ab}\sum_{I_a=1}^{2s_a}
 \int_{\tau_{a,I_a-1}}^{\tau_{a,I_a}}dt_{a,I_a}&&\left[\dot{\boldsymbol
r}_{a,I_a}(t_{a,I_a})\cdot\boldsymbol B^{ab}\left(\boldsymbol
r_{a,I_a}(t_{a,I_a}),t_{a,I_a}\right)\right.\nonumber\\
&&\left. +B^{ab}_3\left(\boldsymbol
r_{a,I_a}(t_{a,I_a}),t_{a,I_a}\right)\right] \nonumber\\
+\frac{i\kappa}{8\pi^2}\sum_{a=1}^{N-1}\sum_{b=a+1}^N
\sum_{I_b=1}^{2s_b}
 \int_{\tau_{b,I_b-1}}^{\tau_{b,I_b}}dt_{b,I_b}&&\left[\dot{\boldsymbol
r}_{b,I_b}(t_{b,I_b})\cdot\boldsymbol C^{ab}\left(\boldsymbol
r_{b,I_b}(t_{b,I_b}),t_{b,I_b}\right)\right.\nonumber\\
&&\left. +C^{ab}_3\left(\boldsymbol
r_{b,I_b}(t_{b,I_b}),t_{b,I_b}\right)\right].\label{stop}
\end{eqnarray} 
After going back to the parametrization of the loops $\Gamma_a$ with the help of the arc-lengths using 
Eqs.~(\ref{integrali}) and (\ref{integralinewpar}) and integrating out the BF fields, it is possible to recover 
in the expression of ${\cal Z}_{\rm pol}(\{\lambda\})$ the factors 
$\prod_{a=1}^{N-1}\prod_{b=a+1}^{N}{\rm e}^{+i\lambda_{ab}\chi(\Gamma_a,\Gamma_b)}$
that originate from the Fourier representation of the Dirac delta functions 
$\prod_{a=1}^{N-1}\prod_{b=a+1}^{N}\delta\left(m_{ab}-\chi(\Gamma_a,\Gamma_b)\right)$.
The integration over the BF fields in ${\cal Z}_{\rm pol}(\{\lambda\})$ 
can be performed applying the formula:
\begin{equation}
\int \prod_{a=1}^{N-1}\prod_{b=a+1}^{N}{\cal D}B_\mu^{ab}(x){\cal D}C_\mu^{ab}(x)
{\rm e}^{- i (S_{\rm BF}+S_{\rm top})}=\prod_{a=1}^{N-1}\prod_{b=a+1}^{N}{\rm e}^{+i\lambda_{ab}\chi(\Gamma_a,\Gamma_b)}.
\end{equation}
Let us note that in the above equation the gauge fields have been
quantized using the 
covariant Lorentz gauge. 

\subsection{Passage to Field Theory II: the non-topological interactions}
Analogously to what has been done in the case of the topological interactions, also the interaction terms in 
$S_{\rm EV}$ can be made linear and  local with the help of auxiliary fields. The strategy to achieve this goal 
is a straightforward generalization of that followed by de Gennes and 
co-workers in Refs.~\cite{degennes}. 

For our purposes, it will be convenient to introduce the set of real scalar fields $\varphi_{a,I_a}$, $a=1,\ldots,N$ and $I_a=1,\ldots,2s_a$.
The action of these fields is
\begin{equation}
S_{\varphi}[J]=S_{\varphi}[0]+i\int d^3x\varphi_{a,I_a}(x)J^{a,I_a}(x)
\end{equation}
where (here we use the convention that repeated upper and lower indices are summed):
\begin{equation}
S_{\varphi}[0]=\int d^3xd^3y\left[\varphi_{a,I_a}(x)\varphi_{b,I_b}(y)
\tilde{V}^{-1}(x-y)
({\cal M}^{-1})^{a,I_a;b,I_b}\right]
\end{equation}
\begin{equation}
\tilde{V}^{-1}(x-y)=V^{-1}(\boldsymbol x-\boldsymbol y)\delta(x^3-y^3)
\end{equation}
and 
\begin{equation}
\int d^2 \boldsymbol y V(\boldsymbol x-\boldsymbol y)V^{-1}(\boldsymbol y-\boldsymbol z)=\delta(\boldsymbol x-\boldsymbol z).
\end{equation}
In other words, $V^{-1}(\boldsymbol x-\boldsymbol y)$ is the operator that inverts the potential $V(\boldsymbol r)$ appearing in $S_{\rm EV}$.
The currents $J^{a,I_a}(x)$ are defined as follows
\begin{equation}
J^{a,I_a}(x)=\int_{\tau_{a,I_a-1}}^{\tau_{a,I_a}}dt_{a,I_a}
\delta^{(2)}(\boldsymbol x-\boldsymbol r_{a,I_a}(t_{a,I_a}))
\delta(x^3-t_{a,I_a})(-1)^{I_a-1}.
\end{equation}
${\cal M}^{-1}$ is the inverse of the matrix 
(we consider $a,I_a$ and $b,I_b$ as composite indexes denoting
respectively the rows and columns) defined in Eq.~(\ref{matrixm}). 

Supposing that $\cal M$ is a $n\times n-$dimensional matrix, it is
easy to find its inverse, which is given by: 
\begin{equation}
{\cal M}^{-1}=\left(
{\begin{array}{cccc}
\frac{n-2}{n-1}&-\frac{1}{n-1}&\ldots&-\frac{1}{n-1}\\
-\frac{1}{n-1}&\frac{n-2}{n-1}&\ldots&-\frac{1}{n-1}\\
\vdots&\vdots&\ddots&\vdots\\
-\frac{1}{n-1}&-\frac{1}{n-1}&\ldots&\frac{n-2}{n-1}
\end{array}}
\right)
\end{equation}
In words, ${\cal M}^{-1}$ is the matrix whose diagonal elements are
$\frac{n-2}{n-1}$, while all the other elements are
$-\frac{1}{n-1}$. Let us note  
that in the present case $n=N(s_1+s_2+\ldots +s_N)$.
It is possible to show that, apart from an irrelevant constant
\begin{equation}
\int \prod_{a=1}^{N}\prod_{I_a=1}^{2s_a}
{\cal D}\varphi_{a,I_a}{\rm e}^{-S_\varphi[J]}={\rm e}^{-S_{\rm EV}} \label{idetwo}
\end{equation}
where $S_{\rm EV}$ is written in the form of Eq.~(\ref{SEV}).

\subsection{Passage to Field Theory III: Second quantization}
Putting all together, the probability function ${\cal Z}_{\rm pol}(\{\lambda\})$ 
of Eq.~(\ref{ggg}) may be expressed in terms of the auxiliary fields 
$B_{\mu}^{ab}(x)$, $C_{\mu}^{ab}(x)$ and $\varphi_{a,I_a}(x)$ as follows
\begin{equation}
{\cal Z}_{\rm pol}(\{\lambda\})=
\int {\cal D}(fields)\,{\rm e}^{-iS_{\rm BF}}
\,{\rm e}^{-S_{\varphi}[0]}\prod_{a=1}^{N}\prod_{I_a=1}^{2s_a}\int {\cal D}\boldsymbol 
r_{a,I_a}(t_{a,I_a})\,{\rm e}^{-S_{\rm part}(\boldsymbol r_{a,I_a})} \label{zpol3}
\end{equation}
where each of the actions $S_{\rm part}(\boldsymbol r_{a,I_a})$,
$a=1,\ldots,N$ and $I_a=1,\ldots,2s_a$, formally coincides 
with the action of a particle immersed in the external potential 
$\varphi_{a,I_a}(t_{a,I_a})$ and in an external magnetic field that consists in a linear combination of the fields $B_{\mu}^{ab}$ and
 $C_{\mu}^{ab}$:
\begin{eqnarray}
S_{\rm part}(\boldsymbol r_{a,I_a})&=&\int_{\tau_{a,I_a-1}}^{\tau_{a,I_a}}
dt_{a,I_a}\left[(-1)^{I_a-1}g_{a,I_a}\dot{\boldsymbol r}_{a,I_a}^2
(t_{a,I_a})+i\varphi_{a,I_a}(\boldsymbol r_{a,I_a}(t_{a,I_a}),t_{a,I_a})(
-1)^{I_a-1}\right.\nonumber\\
&&\left.+i\dot{\boldsymbol r}_{a,I_a}
(t_{a,I_a})\cdot\boldsymbol A^a(\boldsymbol
r_{a,I_a}(t_{a,I_a}),t_{a,I_a})+iA^a_3(\boldsymbol
r_{a,I_a}(t_{a,I_a}),t_{a,I_a})\right]. \label{spol} 
\end{eqnarray}
In Eq.~(\ref{spol}) we have put 
\begin{equation}
A_\mu^1(\boldsymbol r,t)=\sum_{b=2}^N\lambda_{1b}B_\mu^{1b}(\boldsymbol r,t)\label{a11}
\end{equation}
\begin{equation}
A_\mu^a(\boldsymbol
r,t)=\sum_{b=a+1}^N\lambda_{ab}B_\mu^{ab}(\boldsymbol r,t)+
\frac{\kappa}{8\pi^2}\sum_{c=1}^{a-1}C_\mu^{ca}(\boldsymbol r,t)\qquad
a=2,\ldots,N-1 
\end{equation}
\begin{equation}
A_\mu^N(\boldsymbol r,t)=\frac{\kappa}{8\pi^2}\sum_{c=1}^{N-1}C_\mu^{cN}(\boldsymbol r,t)\label{a33}
\end{equation}
and
\begin{equation}
{\cal D}(fields)=\left[\prod_{a=1}^{N-1}\prod_{b=a+1}^{N}\int{\cal D}B_\mu^{ab}C_\mu^{ab}\right]\left[\prod_{a=1}^{N}\prod_{I_a=1}^{2s_a}\int{\cal D}\varphi_{a,I_a}\right].
\end{equation}
Let us note that with Eq.~({\ref{zpol3}}) we have succeeded to rewrite
the probability function  
${\cal Z}_{\rm pol}(\{\lambda\})$ in such a way that
the subtrajectories $\boldsymbol r_{a,I_a}(t_{a,I_a})$ do not interact
directly with each other. They interact only indirectly via the fields
$\varphi_{a,I_a}$ and $A_\mu^a$. 

The problem of passing to second quantized path integral in the case of a particle with partition function:
\begin{equation}
{\cal Z}_{\rm part}^{a,I_a}=\int{\cal D}\boldsymbol r_{a,I_a}(t_{a,I_a})\,
{\rm e}^{-S_{\rm part}(\boldsymbol r_{a,I_a})}
\end{equation}
is very well known in polymer
physics~\cite{FELAPLB1998,FELAJPA1999,FFNOVA,FFAnnPhys2002}. After
introducing $n_{a,I_a}-$multiplets of complex replica fields: 
\begin{eqnarray}
\vec{\Psi}(\boldsymbol x,t)&=&
(\psi^1_{a,I_a}(\boldsymbol x,t),\ldots,\psi^{n_{a,I_a}}_{a,I_a}(\boldsymbol x,t))\label{mult1}\\
\vec{\Psi}^{\ast}(\boldsymbol x,t)&=&
(\psi^{1\ast}_{a,I_a}(\boldsymbol
x,t),\ldots,\psi^{\ast n_{a,I_a}}_{a,I_a}(\boldsymbol
x,t))\label{mult2} 
\end{eqnarray}
we obtain
\begin{eqnarray}
{\cal Z}_{\rm part}^{a,I_a}=\lim_{n_{a,I_a}\rightarrow 0}\int{\cal
  D}\vec{\Psi}_{a,I_a}{\cal
  D}\vec{\Psi}_{a,I_a}^{\ast}\psi_{a,I_a}^{1\ast}(\boldsymbol
r_{a,I_a}(\tau_{a,I_a}),\tau_{a,I_a}) 
\nonumber\\
\psi_{a,I_a}^{1}(\boldsymbol r_{a,I_a}(\tau_{a,I_a-1}),\tau_{a,I_a-1})\,
{\rm e}^{-S_{\rm part}(\vec{\Psi}_{a,I_a}^{\ast},\vec{\Psi}_{a,I_a})}\label{zpartII}
\end{eqnarray}
where
\begin{eqnarray}
S_{\rm part}(\vec{\Psi}_{a,I_a}^{\ast},
\vec{\Psi}_{a,I_a})&=&\int_{\tau_{a,I_a-1}}^{\tau_{a,I_a}}
dt_{a,I_a}\int 
d^2 \boldsymbol x\left[\vec{\Psi}_{a,I_a}^{\ast}
\frac{\partial}{\partial t}\vec{\Psi}_{a,I_a}
\right.\nonumber\\
&&\left.+\frac{1}{4g_{a,I_a}}\left|\left(\boldsymbol
\nabla-i(-1)^{I_a-1}\boldsymbol A^a\right)\vec{\Psi}_{a,I_a}\right|^2 
\right.\nonumber\\
&&\left.
+\;i\left|\vec{\Psi}_{a,I_a}\right|^2\left(A^a_3
+\;\varphi_{a,I_a}(-1)^{I_a-1}\right)\right]. \label{spart}
\end{eqnarray}
In writing Eq.~(\ref{spart}) and in all the
formulas below 
we  follow the convention that, whenever 
products of $\vec{\Psi}_{a,I_a}^*$ with 
$\vec{\Psi}_{a,I_a}$ appear, also
 the scalar product over the
replica multiplets is implicitly understood.

Eventually, the probability function ${\cal Z}_{\rm pol}(\{\lambda\})$
of Eq.~(\ref{zpol3}) becomes
\begin{equation}
{\cal Z}_{\rm pol}(\{\lambda\})=\int{\cal D}(fields)\,{\rm e}^{-iS_{\rm BF}}
\,{\rm e}^{-S_{\varphi}[0]}\prod_{a=1}^{N}\prod_{I_a=1}^{2s_a}{\cal Z}_{\rm part}^{a,I_a}
\end{equation}
with ${\cal Z}_{\rm part}^{a,I_a}$ given by Eq.~(\ref{zpartII}).
From the actions $S_{\rm part}(\vec{\Psi}_{a,I_a}^{\ast},
\vec{\Psi}_{a,I_a})$ 
shown in Eq.~(\ref{spart}), 
we see that the topological forces are tightly related to
the non-topological forces mediated by the potential
$V(\boldsymbol x-\boldsymbol y)$.
This can be realized from the fact that the fields $\varphi_{a,I_a}$ and
the third component of the vector fields $A_3^a$ are coupled in the
same way 
with the matter fields $\vec{\Psi}_{a,I_a}$ and 
$\vec{\Psi}_{a,I_a}^{\ast}$.
This interplay between topological and non-topological interactions remains
 explicit after the integration over the auxiliary $\varphi_{a,I_a}$.
After performing these integrations, we arrive at the final expression 
of ${\cal Z}_{\rm pol}(\{\lambda\})$:
\begin{eqnarray}
{\cal Z}_{\rm pol}(\{\lambda\})&=&\left[\prod_{c=1}^{N-1}\prod_{d=c+1}^{N}\int{\cal D}B_{\mu}^{cd}{\cal D}C_{\mu}^{cd}\right]\nonumber\\
&&\left[\prod_{a=1}^{N}\prod_{I_a=1}^{2s_a}
\lim_{n_{a,I_a}\rightarrow 0}\int{\cal D}\vec{\Psi}_{a,I_a}^{\ast}{\cal D}\vec{\Psi}_{a,I_a}
\psi_{a,I_a}^{1\ast}(\boldsymbol r_{a,I_a}(\tau_{a,I_a}),
\tau_{a,I_a})\right.\nonumber\\
&&\left.
\psi_{a,I_a}^1(\boldsymbol r_{a,I_a}(\tau_{a,I_a-1}),\tau_{a,I_a-1})
\right]
{\rm e}^{-iS_{\rm BF}}\,
{\rm e}^{-S_{\rm matter}}\label{zpolfinal}
\end{eqnarray}
where $S_{\rm BF}$ has been already defined in Eq.~(\ref{ppm}) and
\begin{equation}
S_{\rm matter}=S_{\rm matter}^1+S_{\rm matter}^2\label{smatter}
\end{equation}
with
\begin{eqnarray}
S_{\rm matter}^1=\sum_{a=1}^N\sum_{I_a=1}^{2s_a}\int_{\tau_{a,I_a-1}}^{\tau_{a,I_a}}dt_{a,I_a}\int
d^2\boldsymbol x 
\left[
\vec{\Psi}_{a,I_a}^{\ast}\left(\frac{\partial}{\partial t}+
i A_3^a\right)\vec{\Psi}_{a,I_a}\right.\nonumber\\
\left.
+\frac{1}{4g_{a,I_a}}\left|\left(\boldsymbol 
\nabla-i(-1)^{I_a-1}\boldsymbol A^a\right)\vec{\Psi}_{a,I_a}
\right|^2
\right]\label{smatter1}
\end{eqnarray}
and
\begin{eqnarray}
S_{\rm matter}^2 &=& \sum_{a,b=1}^N\sum_{I_a=1}^{2s_a}\sum_{I_b=1}^{2s_b}
\int_{\tau_{a,I_a-1}}^{\tau_{a,I_a}}dt_{a,I_a}\int d^2\boldsymbol x
d^2\boldsymbol y\nonumber
\\
&\times&
\frac{\cal M}{4}^{a,I_a;b,I_b}
\left|\vec{\Psi}_{a,I_a}(\boldsymbol x,t)
\right|^2 V(\boldsymbol x-\boldsymbol y) 
\left|\vec{\Psi}_{b,I_b}(\boldsymbol y,t)
\right|^2.\label{smatter2}
\end{eqnarray}
Looking at Eqs.~\ceq{zpolfinal}--\ceq{smatter2}, we see that  
the original polymer partition function \ceq{ppm} has been transformed
into a  field theory of two-dimensional quasiparticles. 
The action $S_{\rm matter}^1$ in Eq.~(\ref{smatter1}) is formally
equivalent to the action of a 
multicomponent 
system of anyons subjected to the interactions described by
the action $S^2_{\rm matter}$ in Eq.~(\ref{smatter2}). 
Similar systems have been discussed in connection with the
fractional quantum Hall effect and high $T_{C}$ superconductivity
\cite{wilczek2}.  
The only differences in our case are the boundaries of the
integrations over the 
time, which in this work
 depend on the heights of the points of maxima and
minima of the two trajectories $\Gamma_1,\ldots,\Gamma_N$. Moreover,
here the quasiparticles
are bosons of spin $n_{a,I_a}$, $a=1,\ldots,N$ and $I_a=1,\ldots,2s_a$
considered in the limit 
$n_{a,I_a}\rightarrow 0$. 

At this point, we quantize the BF fields using the Coulomb gauge 
and perform the integration over the third components $B_3^{ab}$ 
and $C_3^{ab}$. The generalization of Eq.~(\ref{bcaction}) to the
case of $N$ loops $\Gamma_1,\ldots,\Gamma_N$ is straightforward. The
BF action $S_{\rm BF}$ becomes in the Coulomb gauge: 
\begin{equation}
S_{\rm BF}=\sum_{a=1}^{N-1}\sum_{b=a+1}^N\frac{\kappa}{4\pi}
\int d^2\boldsymbol xdt\left[B_3^{ab}\Delta c^{ab}+C_3^{ab}\Delta
  b^{ab}\right] 
\label{sbfcgN}
\end{equation}
where $b^{ab}$ and $c^{ab}$ are scalar fields related to the Hodge 
decomposition (\ref{hodgedef}).
The third components of the BF fields play the role of Lagrange 
multipliers. 
They can be easily integrated out in the probability function ${\cal
  Z}_{\rm pol}(\{\lambda\})$ of Eq.~(\ref{zpolfinal}). As a result of this
operation,  the
following constraints are imposed: 
\begin{eqnarray}
\frac{\kappa}{4\pi}\Delta c^{ab}+\lambda_{ab}\sum_{I_a=1}^{2s_a}
|\vec{\Psi}_{a,I_a}|^{2}
\theta(\tau_{a,I_a}-t)\theta(t-\tau_{a,I_a-1})&=&0 \;\;\;
\left\{ \begin{array}{rcl} a&=&1,\ldots,N-1 \\
 b&=&2,\ldots,N \end{array}\right.\label{constr1}\\
\Delta b^{ab}+\frac{1}{2\pi}\sum_{I_b=1}^{2s_b}
|\vec{\Psi}_{b,I_b}|^{2}
\theta(\tau_{b,I_b}-t)\theta(t-\tau_{b,I_b-1})&=&0 \;\;\;
\left\{\begin{array}{rcl} b&=&2,\ldots,N \\
a&=&1,\ldots,b-1\end{array}\right. \label{constr4} 
\end{eqnarray}
The final form of the probability function ${\cal Z}_{\rm pol}(\{\lambda\})$
in the Coulomb gauge is
\begin{eqnarray}
{\cal Z}_{\rm pol}(\{\lambda\})&=&\left[\prod_{c=1}^{N-1}\prod_{d=c+1}^{N}\int{\cal D}\boldsymbol B^{cd}{\cal D}
\boldsymbol C^{cd}\right]\nonumber\\
&&\hspace{-60pt}\left[\prod_{a=1}^{N}\prod_{I_a=1}^{2s_a}
\lim_{n_{a,I_a\rightarrow 0}}\int {\cal D}\vec{\Psi}^\ast_{a,I_a}
{\cal D}\vec{\Psi}_{a,I_a}\psi_{a,I_a}^{1\ast}(\boldsymbol r_{a,I_a}
(\tau_{a,I_a}),\tau_{a,I_a})\psi_{a,I_a}^{1}(\boldsymbol r_{a,I_a}
(\tau_{a,I_a}),\tau_{a,I_a})
\right]\nonumber\\
&\times& {\rm e}^{-S_{\rm matter,CG}}\label{reallyfinal}
\end{eqnarray}
where
\begin{equation}
S_{\rm matter,CG}=S_{\rm matter,CG}^1+S_{\rm matter}^2.\label{reallyfinalS}
\end{equation}
Here $S_{\rm matter}^2$ is the same of Eq.~(\ref{smatter2}) while
\begin{eqnarray}
S_{\rm matter,CG}^1=\sum_{a=1}^{N}\sum_{I_a=1}^{2s_a}\int_{\tau_{a,I_a-1}}
^{\tau_{a,I_a}}dt_{a,I_a}\int d^2\boldsymbol x\frac{1}{4g_{a,I_a}}\left[
|\boldsymbol\nabla\vec{\Psi}_{a,I_a}|^{2}\right.\nonumber\\
\left.+i(-1)^{I_a-1}
\boldsymbol A^a\cdot \boldsymbol{J}^a+|\vec{\Psi}_{a,I_a}|^{2}
(\boldsymbol A^a)^2
\right]. \label{s1matterCG}
\end{eqnarray}
In the above equation the $\boldsymbol{J}^a$'s are the currents
\begin{equation}
\boldsymbol{J}^a=\vec{\Psi}_{a,I_a}\boldsymbol \nabla 
\vec{\Psi}_{a,I_a}^\ast-\vec{\Psi}_{a,I_a}^\ast\boldsymbol \nabla 
\vec{\Psi}_{a,I_a}.
\end{equation}
The BF-fields cease to be independent degrees of freedom because, thanks
to the constraints (\ref{constr1})--(\ref{constr4}), they can be expressed
as functions of the matter fields
$\vec{\Psi}_{a,I_a}^\ast$, $\vec{\Psi}_{a,I_a}$.
As a matter of fact, these constraints can be solved analytically with
respect to the remnants $b^{ab}$, $c^{ab}$ of the original gauge
fields. Remembering that $B_i^{ab}=\epsilon_{ij}\partial^jb^{ab}$ and
 $C_i^{ab}=\epsilon_{ij}\partial^jc^{ab}$, we write down directly the
components of the fields $\boldsymbol B^{ab}$ and $\boldsymbol C^{ab}$:
\begin{eqnarray}
C_i^{ab}(\boldsymbol x,t) &=& -\frac{2\lambda_{ab}}{\kappa}\int d^2\boldsymbol y 
\sum_{I_a=1}^{2s_a}
|\vec{\Psi}_{a,I_a}(\boldsymbol y,t)|^{2}
\epsilon_{ij}\frac{(x-y)^j}{|\boldsymbol x-\boldsymbol y|^2}
\theta(\tau_{a,I_a}-t)\theta(t-\tau_{a,I_a-1})\nonumber
\\
&=&0,\quad\quad\quad\quad
a=1,\ldots,N-1,\quad
b=2,\ldots,N, \label{C1C}
\end{eqnarray}
\begin{eqnarray}
B_i^{ab}(\boldsymbol x,t) &=& -\int d^2\boldsymbol y \frac{1}{4\pi^2}
\epsilon_{ij}\frac{(x-y)^j}{|\boldsymbol x-\boldsymbol y|^2}
\sum_{I_b=1}^{2s_b}
|\vec{\Psi}_{b,I_b}(\boldsymbol y,t)|^{2}
\theta(\tau_{b,I_b}-t)\theta(t-\tau_{b,I_b-1})\nonumber\\
&=& 0,\quad\quad\quad\quad
 b=2,\ldots,N, \quad
 a=1,\ldots,b-1.  \label{C4C}
\end{eqnarray}
The above expressions of the BF-field should be inserted
in Eqs.~(\ref{a11})--(\ref{a33}) which define the fields $\boldsymbol
A^a$ 
appearing in the action~(\ref{smatter1}).
Let us note that the fields $\boldsymbol A^a$ written in terms of the
solutions (\ref{C1C})--(\ref{C4C}) 
 do not contain the parameter $\kappa$ as expected.
Putting all together, it is possible to conclude that the total energy
density of the system of plats contains quartic and sextic 
interactions in the matter fields $\vec{\Psi}_{a,I_a}^\ast$, $\vec{\Psi}_{a,I_a}$.
This conclusion is in agreement with previous calculations performed in 
\cite{ferrari}, where it has been shown that the topological constraints
generate quartic and sextic corrections due to the presence of the 
topological constraints. The difference is that in \cite{ferrari}
the approximate method of the effective potential has been used, while
the present calculations are exact.
\section{A statistical model of a $2s-$plat composed by $N-$ linked
  polymers }\label{sectionVI} 
\begin{figure}[bpht]
\centering
\includegraphics[scale=.5]{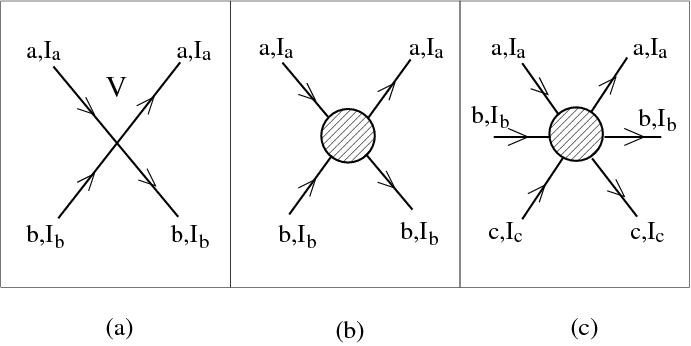}\\
\caption{Feynman diagram representation of the interactions in
  Eqs.~(\ref{smatter2}) and (\ref{s1matterCG}).}\label{vertices} 
\end{figure}

Using Feynman diagrams, the nontopological quartic interactions in 
Eq.~(\ref{smatter2}) may be represented by the four-vertex in
Fig.~\ref{vertices}-(a).
The quartic interactions of topological origin described by the
contributions to $S^1_{\rm matter, CG}$ of Eq.~(\ref{s1matterCG}) in which
the fields $\boldsymbol A^a$ are coupled to the  currents
$\boldsymbol J^a$, correspond to the four-vertex of
Fig.~\ref{vertices}-(b). The sextic interactions, also of topological
origin, 
consisting in the terms in $S^1_{\rm matter,CG}$  proportional to
$(\boldsymbol A^a)^2$, are displayed in Fig.~\ref{vertices}-(c). Let us
note that in both the four-vertex and the six-vertex of
Figs.~\ref{vertices}-(b)
and \ref{vertices}-(c) the external legs depart from a solid
circle. This circle symbolizes the fact that these vertices contain
non-perturbative contributions coming from the path integral summation
over the field $B_\mu^{ab}$  and $C_\mu^{ab}$.
The strengths $g_4$ and $g_6$ of the quartic and sextic interactions
of topological 
origin are respectively proportional to:
\begin{equation}
g_4\sim\frac{\lambda_{ab}}{8\pi^2}\qquad\qquad
g_6\sim\frac{\lambda_{ab}\lambda_{ac}}{16\pi^4} \label{couplings}
\end{equation}
As it is clear from Eq.~(\ref{deltaft}), the $\lambda_{ab}$'s are
Fourier coefficients varying in the interval $(-\infty,+\infty)$. For
this reason, $g_4$ and $g_6$ cannot be considered as real coupling
constant. However, the parameters $\lambda_{ab}$ may be interpreted as
chemical potentials that specify how easy is the linking of two trajectories
$\Gamma_a$ and $\Gamma_b$. To small values of $\lambda_{ab}$
correspond big values of the linking number $m_{ab}$ and viceversa.

\begin{figure}[bpht]
\centering
\includegraphics[scale=.5]{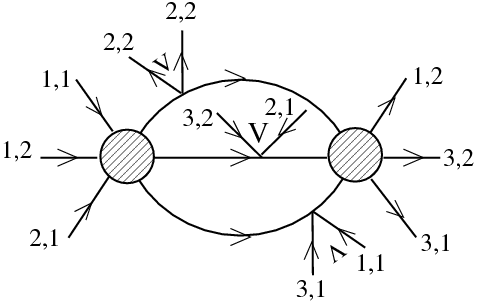}\\
\caption{Example of a process in which the three body interactions of
  topological origin do not vanish in the zero replicas limit $n_{a,I_a}\to0$
  appearing in Eq.~(\ref{reallyfinal}). The process describes the
  interaction of the subtrajectories $\Gamma_{a,I_a}$, $a=1,2,3$,
  $I_a=1,2$ forming a $6-$plat in which three loops
  $\Gamma_1,\Gamma_2,\Gamma_3$ are linked
  together.}\label{three-body-forces-example}   
\end{figure}

An important feature of the model described in
Eqs.~(\ref{reallyfinal}) and (\ref{reallyfinalS}) is that the
interactions of topological origin  have sextic interactions, in which
the monomers of three different loops are involved. The appearance of
three-body forces was up to now not supposed to be possible in the
case of topological constraints imposed using the Gauss linking
number. As a matter of fact, this link invariant controls only the
linking between pairs of polymer rings. In the case $N=2$, in which we have
just two loops, these three-body interactions are suppressed as showed
in Ref.~\cite{ferrari}, because they vanish when the limit in which the
numbers of replicas $n_{a,I_a}$ approach zero is performed in  the
probability function of Eq.~(\ref{reallyfinal}). However, not all diagrams
with three-body interactions disappear when $N>2$. An example of
nontrivial contribution in which interactions of three monomers are
taking place is shown in Fig.~\ref{three-body-forces-example}.

Another characteristic of the model describing the statistical
mechanics of $2s-$plats introduced here is the existence of vortex
solutions of the 
equations that minimize the energy of the static field
configurations. An example of such solutions will be presented in the
next Section in the case $N=2$.
\section{Self-dual solutions of the two-polymer
  problem}\label{sectionVII}
In this Section we restrict ourselves for simplicity to 
$4-$plats. Moreover, the non-topological interactions contained in
$S^2_{\rm matter}$ will be ignored. 
We will also suppose that the replica numbers
are independent of
$I_a$, i.e.:
\begin{eqnarray}
\vec{\Psi}(\boldsymbol x,t)=(\psi^1_{a,I_a}(\boldsymbol
x,t),\ldots,\psi^{n_{a}}_{a,I_a}(\boldsymbol x,t))\qquad a=1,2\mbox{
  and }I_a=1,2\\ 
\vec{\Psi}^{\ast}(\boldsymbol x,t)=(\psi^{1\ast}_{a,I_a}(\boldsymbol x,t),\ldots,\psi^{\ast n_{a}}_{a,I_a}(\boldsymbol x,t))\qquad a=1,2\mbox{
  and }I_a=1,2.
\end{eqnarray} 
In Eqs.~(\ref{mult1}) and (\ref{mult2}) each pair of complex fields
$\vec{\Psi}^*_{a,I_a},\vec{\Psi}_{a,I_a}$ had a separate replica index
$n_{a,I_a}$, but it is easy to check that $I_a-$independent replica
indexes are possible too without jeopardizing the passage to field
theory and in  particular the calculations made in
Section~\ref{sectionIV}.
The partition function of a $4-$plat
formed by two 
linked 
polymers is obtained 
by putting $N=2$ and $s_{1}=s_{2}=1$ in the
general
 partition function of a $2s-$plat given in
Eq.~\ceq{reallyfinal}. Accordingly, the action $S_{\rm matter,CG}$ in
Eq.~\ceq{reallyfinalS} in this particular case becomes 
\begin{eqnarray}
S_{\rm matter,CG}&=&\int_{\tau_{1,0}}^{\tau_{1,1}}\!dt
\int\!d^{2}\boldsymbol x
\Bigg\lbrace\vec{\Psi}_{1,1}^{*}
\left[\frac{\partial}{\partial t}-\frac{1}{4g_{1,1}}\boldsymbol
  D^{2}\left(-\lambda_{12},\boldsymbol B^{12}\right)\right]\vec{\Psi}_{1,1}\nonumber\\ 
&+&\vec{\Psi}_{1,2}^{*}\left[\frac{\partial}{\partial
    t}-\frac{1}{4g_{1,2}}
\boldsymbol D^{2}\left(\lambda_{12},\boldsymbol B^{12}\right)
\right]\vec{\Psi}_{1,2}\Bigg\rbrace
\nonumber \\
&+&
\int_{\tau_{2,0}}^{\tau_{2,1}}\!dt\int d^2\boldsymbol x
\vec{\Psi}_{2,1}^{*}
\Bigg\lbrace\left[
\frac{\partial}{\partial t}
-\frac{1}{4g_{2,1}}
\boldsymbol D^{2}\left(
-\frac{\kappa}{8\pi^2}
,\boldsymbol C^{12}\right)
\right]\vec{\Psi}_{2,1}
\nonumber \\
&+&
\vec{\Psi}_{2,2}^{*}
\left[
\frac{\partial}{\partial t}
-\frac{1}{4g_{2,2}}
\boldsymbol D^{2}
\left(\frac{\kappa}{8\pi^2},\boldsymbol C^{12}\right)
\right]
\vec{\Psi}_{2,2}\Bigg\rbrace. \label{afourplat}
\end{eqnarray}
In the above equation $\boldsymbol D$ denotes the covariant
derivatives, which are of two types depending if they are defined with
respect to the field $\boldsymbol B^{12}$
  or to the field $\boldsymbol C^{12}$:
\begin{equation}
\boldsymbol D(\pm \lambda_{12},\boldsymbol B^{12})=\boldsymbol \nabla \pm
i\lambda_{12} \boldsymbol B^{12}  
\qquad\qquad
\boldsymbol D\left(\pm \frac{\kappa}{8\pi^2},\boldsymbol
C^{12}\right)=\boldsymbol \nabla \pm 
i\frac{\kappa}{8\pi^2} \boldsymbol C^{12}.
\end{equation}
 As mentioned at the end of the
  previous Section, the fields $\boldsymbol B^{12}$
  and $\boldsymbol C^{12}$   are not independent degrees of
  freedom, because they are fully determined by the constraints
  (\ref{constr1})--(\ref{constr4}). In the present case $N=2$, $s_1=s_2=2$,
the required
  conditions are:
\begin{eqnarray}
\epsilon^{ij}\partial_i B_j^{12}&=&-\frac 1{2\pi}\left(
|\vec{\Psi}_{21}|^2+|\vec{\Psi}_{22}|^2
\right)\theta(\tau_{2,1}-t)\theta(t-\tau_{2,0})\label{cc1}\\
\epsilon^{ij}\partial_i C_j^{12}&=&-\frac {4\pi\lambda_{12}}{\kappa}\left(
|\vec{\Psi}_{11}|^2+|\vec{\Psi}_{12}|^2
\right)\theta(\tau_{1,1}-t)\theta(t-\tau_{1,0}).\label{cc2}
\end{eqnarray}
We will consider now the static field configurations that minimize the
action $S_{\rm matter,CG}$ of Eq.~(\ref{afourplat}). From
Ref.~\cite{FFPLA2004}
it is known that this action admits self-dual solutions in the case in
which the parameters $g_{a,I_a}$, $a=1,2$ and $I_a=1,2$ are all equal.
To this purpose,  for any constant $\gamma$
and gauge field $\boldsymbol a $ we define the new covariant derivatives
$
D_{\pm}(\gamma,\boldsymbol a)$: 
\begin{equation}
 D_\pm(\gamma,\boldsymbol a)=D_1(\gamma,\boldsymbol a)\pm iD_2
 (\gamma,\boldsymbol a) 
\end{equation}
where $D_1$ and $D_2$ denote the first and second components of the
covariant derivative $\boldsymbol D$. In terms of the $D_\pm$'s, the
self-duality equations may be expressed as follows:
\begin{eqnarray}
 D_{+}\left(-\lambda_{12},\boldsymbol
B^{12}\right)\psi_{1,1}^{n_1}&=&0\label{cleq1}\\ 
D_{+}\left(\lambda_{12},\boldsymbol
B^{12}\right)\psi_{1,2}^{n_1}&=&0\label{cleq2}\\ 
 D_{-}\left(-\frac{\kappa}{8\pi^2},\boldsymbol C^{12}
\right)\psi_{2,1}^{n_2}&=&0 
\label{cleq3}\\
 D_{-}\left(\frac{\kappa}{8\pi^2},\boldsymbol
C^{12}\right)\psi_{2,2}^{n_2}&=&0.\label{cleq4} 
\end{eqnarray}
We notice in the
constraints (\ref{cc1}) and (\ref{cc2}) the cumbersome presence of the
Heaviside $\theta-$functions. They are required in order to take into account
the fact that the heights of the points belonging to the
subtrajectories $\Gamma_{a,I_a}$ are 
only partially overlapping.
As a consequence, to avoid  complications,
 we  will assume that
$\tau_{1,0}=\tau_{2,0}=\tau_0$ and $\tau_{1,1}=\tau_{2,1}=\tau_1$,
i.e. all subtrajectories will start and end at the same height.
In this way the Heaviside $\theta-$functions are no longer needed.
Moreover, 
we will
 restrict ourselves to  replica symmetric solutions
by putting:
\begin{eqnarray}
\psi^1_{1,I_1}=\cdots
=\psi_{1,I_1}^{n_1}&=&\psi_{1,I_1}\quad \mbox{for} \quad  I_1=1,2\nonumber\\
\psi_{2,I_2}^1=\cdots= \psi_{2,I_2}^{n_2}&=&\psi_{2,I_2} \quad
\mbox{for} \quad I_2=1,2.
\end{eqnarray}
After these simplifications, the self-duality conditions
(\ref{cleq1})--(\ref{cleq4}) and the constraints (\ref{cc1}) and
(\ref{cc2}) become:
\begin{eqnarray}
\left[\partial_1-i\lambda_{12} B_1^{12}+i\left(\partial_2-i\lambda_{12} B_2^{12}\right)\right]\psi_{1,1}&=&0
\label{eqexp1}\\
\left[\partial_1+i\lambda_{12} B_1^{12}+i\left(\partial_2+i\lambda_{12} B_2^{12}\right)\right]\psi_{1,2}&=&0
\label{eqexp2}\\
\left[\partial_1-\frac{i\kappa}{8\pi^2}
  C_1^{12}-i\left(\partial_2-\frac{i\kappa}{8\pi^2}
  C_2^{12}\right)\right]\psi_{2,1}&=&0 
\label{eqexp3}\\
\left[\partial_1+\frac{i\kappa}{8\pi^2}
  C_1^{12}-i\left(\partial_2+\frac{i\kappa}{8\pi^2}
  C_2^{12}\right)\right]\psi_{2,2}&=&0 
\label{eqexp4}
\end{eqnarray}
and
\begin{eqnarray}
\epsilon^{ij}\partial_iB_j^{12}&=&-\frac 1{2\pi}n_2\left(|
\psi_{2,1}|^2 +
| \psi_{2,2}|^2\right)\label{cstrexp1} \\
\epsilon^{ij}\partial_{i}C_{j}^{12}&=&-\frac{4n_1\pi \lambda_{12}}{\kappa}
\left(|\psi_{1,1}|^2+|\psi_{1,2}|^2\right).\label{cstrexp2}
\end{eqnarray}
At this point we pass to polar coordinates by performing the
transformations:
\begin{equation}
\psi_{a,I_a}={\rm e}^{i\omega_{a,I_a}}\rho_{a,I_a}^{1/2}.\label{trsf1}
\end{equation}
After the above change of variables in 
Eqs.~(\ref{eqexp1}--\ref{cstrexp2}) and separating the real and
imaginary parts, we obtain: 
\begin{eqnarray}
\partial_1\omega_{1,1}-\lambda_{12}
B_1^{12}+\frac{1}{2}\partial_2\log\rho_{1,1}&=&0 
\label{eqpc1}\\
-\partial_2\omega_{1,1}+\lambda_{12}
B_2^{12}+\frac{1}{2}\partial_1\log\rho_{1,1}&=&0 
\label{eqpc2}\\
\partial_1\omega_{1,2}+\lambda_{12}
B_1^{12}+\frac{1}{2}\partial_2\log\rho_{1,2}&=&0 
\label{eqpc3}\\
-\partial_2\omega_{1,2}-\lambda_{12}
B_2^{12}+\frac{1}{2}\partial_1\log\rho_{1,2}&=&0 
\label{eqpc4}\\
\partial_1\omega_{2,1}-\frac{\kappa}{8\pi^2}
C_1^{12}-\frac{1}{2}\partial_2\log\rho_{2,1}&=&0 
\label{eqpc5}\\
\partial_2\omega_{2,1}-\frac{\kappa}{8\pi^2}
C_2^{12}+\frac{1}{2}\partial_1\log\rho_{2,1}&=&0 
\label{eqpc6}\\
\partial_1\omega_{2,2}+\frac{\kappa}{8\pi^2}
C_1^{12}-\frac{1}{2}\partial_2\log\rho_{2,2}&=&0 
\label{eqpc7}\\
\partial_2\omega_{2,2}+\frac{\kappa}{8\pi^2}
C_2^{12}+\frac{1}{2}\partial_1\log\rho_{2,2}&=&0 
\label{eqpc8}
\end{eqnarray}
\begin{eqnarray}
\epsilon^{ij}\partial_iB_j&=&-\frac 1{2\pi} n_2\left(\rho_{2,1}+
\rho_{2,2}\right)\label{cstrpc1} \\
\epsilon^{ij}\partial_{i}C_{j}&=&-\frac{4n_1\pi \lambda_{12}}{\kappa}
\left(\rho_{1,1}+\rho_{1,2}\right).\label{cstrpc2}
\end{eqnarray}
To solve equations (\ref{eqpc1})--(\ref{eqpc8}) with respect to the
unknowns $\omega_{a,I_a}$ and $\rho_{a,I_a}$, we proceed as follows.
First of all, we isolate from Eq.~(\ref{eqpc1}) and Eq.~(\ref{eqpc3})
the same quantity $\lambda_{12} B_1^{12}$. By requiring
that the expressions of $\lambda_{12} B_1^{12}$ provided by Eqs.~(\ref{eqpc1}) and
(\ref{eqpc3}) are equal, we obtain the consistency condition
\begin{equation}
\partial_1\omega_{1,1}+\frac{1}{2}\partial_2\log\rho_{1,1}=
-\partial_1\omega_{1,2}-\frac{1}{2}\partial_2\log\rho_{1,2}\label{cstc1} 
\end{equation}
A possible solution of Eq.~(\ref{cstc1}) is
\begin{equation}
\omega_{1,1}=-\omega_{1,2}\quad \mbox{and} \quad
\rho_{1,1}=\frac{A_1}{\rho_{1,2}}\label{ansatz1} 
\end{equation}
where $A_1$ is at most a function of $x^1$.
As well, we require that the two different expressions of the
quantity $\lambda_{12} B_2^{12}$ obtained from
Eqs.~(\ref{eqpc2}) and (\ref{eqpc4}) are equal. 
On this way one obtains a condition  analogous to
(\ref{cstc1}),
%
which may be solved by applying the
ansatz (\ref{ansatz1})  and additionally requiring that $A_1$ is a constant.
In a similar way, it is possible to extract from
equations 
(\ref{eqpc5}--\ref{eqpc8}) the conditions:
\begin{equation}
\omega_{2,1}=-\omega_{2,2}\quad \mbox{and} \quad \rho_{2,1}=\frac{A_2}{\rho_{2,2}}\label{ansatz2}
\end{equation}
with $A_2$ being a constant.

Thanks to Eqs.~(\ref{ansatz1}) and (\ref{ansatz2}),
the number of unknowns to be computed is reduced. For instance, if we
choose as independent degrees of freedom
$\omega_{1,1},\omega_{2,1},\rho_{1,1}$ and $\rho_{2,1}$,  the remaining
classical field configurations $\omega_{1,2},\omega_{2,2},\rho_{1,2}$ 
and $\rho_{2,2}$ can be derived using such equations.  
As a consequence, the
system of equations (\ref{eqpc1})--(\ref{cstrpc2}) reduces to:
\begin{eqnarray}
\lambda_{12}
B_1^{12}&=&\partial_1\omega_{1,1}+\frac{1}{2}
\partial_2\log\rho_{1,1}\label{eqred1}\\ 
\lambda_{12}
B_2^{12}&=&\partial_2\omega_{1,1}-\frac{1}{2}
\partial_1\log\rho_{1,1}\label{eqred2}\\
\frac{\kappa}{8\pi^2}
C_1^{12}&=&\partial_1\omega_{2,1}-\frac{1}{2}
\partial_2\log\rho_{2,1}\label{eqred3}\\
\frac{\kappa}{8\pi^2}
C_2^{12}&=&\partial_2\omega_{2,1}+\frac{1}{2}
\partial_1\log\rho_{2,1}\label{eqred4}\\
\partial_1B_2^{12}-\partial_2
B_1^{12}&=&-\frac 1{2\pi}n_2
\left(\rho_{2,1}+\frac{A_2}{\rho_{2,1}}\right)\label{cstrred1}\\
\partial_1C_2^{12}-\partial_2 C_1^{12}&=&-
\frac{4n_1\pi\lambda_{12}}{\kappa}\left(\rho_{1,1}+
\frac{A_1}{\rho_{1,1}}\right)\label{cstrred2}
\end{eqnarray}
where
we have used the fact that
$\epsilon^{ij}\partial_iB_j=\partial_1B_2^{12}-\partial_2B_1^{12}$ and
$\epsilon^{ij}\partial_iC_j=\partial_1C_2^{12}-\partial_2C_1^{12}$.
Eqs.~(\ref{eqred1})--(\ref{cstrred2}) contain
 the unknowns $\omega_{1,1},\omega_{2,1},\rho_{1,1}$ and
$\rho_{2,1}$ that will  be determined below.

By subtracting term by term the two
equations resulting from the derivation of
Eqs.~(\ref{eqred1}) and (\ref{eqred2})
with respect to the variables $x^2$ 
and $x^1$ respectively, we obtain as an upshot the relation:
\begin{equation}
\lambda_{12}\left(\partial_1B_2^{12}-
\partial_2B_1^{12}\right)=\partial_1\partial_2\omega_{1,1}-
\partial_2\partial_1\omega_{1,1}-\frac{1}{2}\Delta\log\rho_{1,1}\label{sub12}
\end{equation}
with $\Delta=\partial_1^2+\partial_2^2$ being the two-dimensional
Laplacian.\\ 
Assuming that $\omega_{1,1}$ is a regular function satisfying the
relation 
\begin{equation}
\partial_1\partial_2\omega_{1,1}-\partial_2\partial_1\omega_{1,1}=0
\end{equation}
Eq.~(\ref{sub12}) becomes:
\begin{equation}
\lambda_{12}\left(\partial_1B_2^{12}-\partial_2B_1^{12}\right)=
-\frac12\Delta\log\rho_{1,1}.\label{sub12nv}
\end{equation}
An analogous identity can be derived starting from Eqs.~(\ref{eqred3})
and (\ref{eqred4}): 
\begin{equation}
\frac{\kappa}{4\pi^2}\left(\partial_1C_2^{12}-\partial_2
C_1^{12}\right)=\Delta\log\rho_{2,1}.\label{sub34nv} 
\end{equation}
The compatibility of (\ref{sub12nv}) and (\ref{sub34nv}) with the
constraints (\ref{cstrred1}) and (\ref{cstrred2}) respectively leads
to the following conditions between $\rho_{1,1}$ and $\rho_{2,1}$:
\begin{eqnarray}
\Delta\log\rho_{1,1}&=&\frac{\lambda_{12}
n_2}\pi\left(\frac{A_2}{\rho_{2,1}}+\rho_{2,1}\right)\label{cnstcy1}\\
\Delta\log\rho_{2,1}&=&-\frac{\lambda_{12} n_1}\pi
\left(\rho_{1,1}+\frac{A_1}{\rho_{1,1}}\right).\label{cnstcy2} 
\end{eqnarray}
The fact that $\rho_{1,1}$ and $\rho_{2,1}$
appear in a symmetric way in Eqs.~(\ref{cnstcy1}) and (\ref{cnstcy2}),
suggests the following ansatz:
\begin{equation}
\rho_{2,1}=\frac{A_3}{\rho_{1,1}}\label{ansatz3}
\end{equation}
with $A_3$ being a constant.
It is easy to check that with this ansatz Eqs.~(\ref{cnstcy1}) and
(\ref{cnstcy2}) remain compatible provided:
\begin{equation}
\frac{A_2}{A_3} = \frac{n_1}{n_2}\quad\mbox{and}\quad\frac{A_3}{A_1}
= \frac{n_1}{n_2} .
\label{constconst}
\end{equation}
We choose $A_1$ to be the independent constant, while $A_2$ and $A_3$
are constrained 
by Eq.~\ceq{constconst} to be dependent on $A_1$:
\begin{equation}
A_2 = \left(\frac {n_1}{n_2}\right )^2 A_1\qquad A_3 = \frac {n_1}{n_2}A_1.
\end{equation}
We are now left only with the task of computing the explicit
expression of $\rho_{1,1}$.
This may be obtained by solving  
the equation:
\begin{equation}
\Delta\log{\rho_{1,1}} = \frac{\lambda_{12} n_1}\pi \left(  \frac{A_1}{\rho_{1,1}}
+\rho_{1,1}   \right) 
\label{rho11fmal}
\end{equation}
The other quantities $\rho_{2,1}$, $\rho_{1,2}$ and $\rho_{2,2}$ can
be derived using  
the relations \ceq{ansatz3}, \ceq{ansatz1} and \ceq{ansatz2}
respectively.
Eq.~(\ref{rho11fmal}) may be cast in a more familiar form by putting:
$\eta=\ln\left(\frac{\rho_{1,1}}{\sqrt{A_1}}\right)$. After this
substitution, Eq.~(\ref{rho11fmal}) becomes the Euclidean cosh--Gordon
equation with respect to $\eta$:
\beq{\Delta\eta=\frac{2\lambda_{12} n_1}\pi\sqrt{A_1}\cosh\eta}{sinhgordon}
Next, it is possible to determine the magnetic fields $\boldsymbol B^{12}$ and
$\boldsymbol C^{12}$ from 
Eqs.~\ceq{cstrred1} and \ceq{cstrred2}. In the Coulomb gauge, in fact,
the two-dimensional 
vector potentials $\boldsymbol B^{12}$ and $\boldsymbol C^{12}$ can be
represented using 
two scalar fields 
$b^{12}$ and $c^{12}$ as follows (see also Eq.~(\ref{hodgedef})):
\begin{equation}
\boldsymbol B^{12} = (-\partial_2 b^{12}, \partial_1 b^{12}) \qquad
\boldsymbol C^{12} = 
(-\partial_2 c^{12}, \partial_1 c^{12}) 
\label{hodgedec}
\end{equation}
Performing the above substitutions
in Eqs.~\ceq{cstrred1} and \ceq{cstrred2}, it turns out that $b^{12}$ and
$c^{12}$ satisfy the relations: 
\begin{equation}
\Delta b^{12} = -\frac{n_1}{2\pi} (\rho_{1,1} + \frac {A_1}{\rho_{1,1}})
\label{eqb}
\end{equation}
\begin{equation}
\Delta c^{12} = -\frac{4n_1\pi\lambda_{12}}{\kappa} (\rho_{1,1} + \frac
{A_1}{\rho_{1,1}}) 
\label{eqc}
\end{equation}
The solution of Eqs.~\ceq{eqb} and \ceq{eqc} can be easily derived
with the help  
of the method of the Green functions once the expression of
$\rho_{1,1}$ is known. 
Finally, the phases $\omega_{1,1}$, $\omega_{1,2}$, $\omega_{2,1}$ and
$\omega_{2,2}$ 
are computed using Eqs.~\ceq{eqred1}--\ceq{eqred4}.
In fact, remembering that we assumed that $\omega_{1,1}=-\omega_{1,2}$
and $\omega_{2,1} = -\omega_{2,2}$ 
in \ceq{ansatz1} and \ceq{ansatz2} respectively, we have only to
determine $\omega_{1,1}$ and $\omega_{2,1}$. 
By deriving Eq.~\ceq{eqred1} with respect to $x^1$ and
Eq.~\ceq{eqred2} with respect to $x^2$, we obtain: 
\begin{eqnarray}
\lambda_{12} \partial_1B_1^{12} &=& \partial_1^2 \omega_{1,1}  + \frac 12
\partial_1\partial_2\log{\rho_{1,1}}\nonumber\\ 
\lambda_{12} \partial_2B_2^{12} &=& \partial_2^2 \omega_{1,1}  - \frac 12
\partial_2\partial_1\log{\rho_{1,1}} 
\end{eqnarray} 
On the other side, by adding term by term the above two equations and
using the fact that 
in the Coulomb gauge
the magnetic field $\boldsymbol B^{12}$ is completely transverse, it is
possible to show that:
\begin{equation}
\Delta \omega_{1,1} = 0 \label{fdsfsd1}
\end{equation}
Proceeding in a similar way with Eq.~\ceq{eqred3} and \ceq{eqred4} it
is possible to derive also the relation 
satisfied by $\omega_{2,1}$:
\begin{equation}
\Delta\omega_{2,1} = 0\label{fdsfsd2}
\end{equation}
\section{Conclusions}\label{sectionVIII}
In this work a $2s-$plat composed by $N$ polymers forming a nontrivial link has
been considered.  The nontrivial interactions and
the topological constraints make the energy density of the system
complicated and nonlocal, but it can be simplified with the
introduction of auxiliary fields. The final model which we obtain is a
standard field theory involving a set of complex scalar fields with 
 sextic interactions  at most. This model
 allows also some phenomenological
predictions that were a priori not obvious and that will be summarized below.
\begin{enumerate}
\item In the case of a $4-$plat, it has been shown  in
  \cite{FFPLA2004} with the help of 
  a Bogomol'nyi tranformation
  that, after eliminating the 
fields  $B_\mu^{ab}$ and $C_\mu^{ab}$, the topological constraints
imposed with the Gauss linking 
  number are responsible for quartic interaction terms in the
  Hamiltonian of the system. In the particular case in which the two-body
  potential of the non-topological interactions is given by
  Eq.~(\ref{twobody}),
these quartic terms are exactly of the
  same form of those contained in the action~(\ref{smatter2}).
Here we have seen in the more general case of a $2s-$plat and an
arbitrary two-body potential $V(\boldsymbol r_2-\boldsymbol r_1)$ how
the interactions arising due to the presence of the constraints
interphere with the  non-topological interactions of Eq.~\ceq{SEV}.
For example, from Eq.~\ceq{spart} it is possible to realize that in the
action $S_{part}(\vec{\Psi}_{a,I_a}^{\ast},
\vec{\Psi}_{a,I_a})$ the third components of the  BF-fields
can be absorbed by 
the fields $\varphi_{a,I_a}$ after a shift. 
Since the
BF-fields are 
related to the topological constraints and the fields $\varphi_{a,I_a}$
are propagating the
non-topological forces,
this  hints to a strong
interplay
between the topological and non-topological interactions.
Let us note that the effect of the forces of topological origin  may result both in a
reciprocal attraction or repulsion between the monomers, while 
the short range two-body potential \ceq{twobody}, corresponding to the
case in which the polymers are immersed in a solution, can only be
attractive if $V_0<0$ 
or repulsive if $V_0>0$.
\item The field theoretical model of polymeric $2s-$plats defined by
  Eqs.~\ceq{reallyfinal}--\ceq{C4C} shows that three-body forces
  become relevant
  in a system
  of $N$ polymers linked together in which the topological constraints are
  imposed by means of the Gauss linking number. These three-body
  forces have been represented in the 
  form of a Feynman diagram in Fig.~\ref{vertices}-(c) and are
  described in Section~\ref{sectionVI}. An example of process in which
  there are interactions between three monomers at once has been shown
  in Fig.~\ref{three-body-forces-example}.
The existence of three-body
  interactions acting on the monomers was not predicted by previous
  calculations. This is probably because only the case $N=2$ has been
  mainly treated so far. When $N=2$, it turns out that the contribution of
  sextic interactions terms in the action of Eq.~\ceq{s1matterCG},
  which are responsible for the presence of the three-body forces,
  vanishes in the zero replica limit. 
Besides, the appearance of three-body forces is not trivial and not
easy to be predicted, because the Gauss linking number involves only
interactions between pairs of monomers.
\item By using the splitting procedure presented in
  Section~\ref{sectionII} and thanks to the introduction of auxiliary
  fields, the problem of the statistical mechanics of a $2s-$plat has
  been mapped into the dynamics of a system in which quasiparticles of
  different kinds are mixed together. In Ref.~\cite{FFPLA2004} it has
  been shown that systems of this type admit vortex solutions. Out of
  the self-duality regime, vortex magnetic lines associated with
  quasi-particles of different kind can repel or attract
  themselves. After a particular choice of the parameters of the
  theory,
in which the coefficients $g_{a,I_a}$, $a=1,2$  and $I_a=1,2$ are all
equal,
a self-dual point is reached in which attractive
and repulsive forces balance themselves and disappear. A similar
phenomenon, but in a different model, has been recently found in
Ref.~\cite{diamantinitrugenberger}.
In this work, the self-dual vortex conformations have been computed
exactly and explicitly up to the solution of a cosh-Gordon equation.
\end{enumerate}

The  topological properties of the
link formed by the $2s-$plat
have been
described here by using the Gauss linking invariant, which is related
to the abelian BF-model of Eq.~(\ref{CSaction}). 
We have seen in Appendix~\ref{appendixB}
how the topological constraints are fixed
 when the BF-model is quantized in the Coulomb gauge. 
In this gauge are counted 
the
 winding 
numbers of all possible pairs of paths formed by 
the subtrajectories $\Gamma_{a,I_a}$ belonging to a loop $\Gamma_a$ 
and the subtrajectories $\Gamma_{b,I_b}$
 belonging to another loop $\Gamma_b$. The sum of all these winding
 numbers is an integer multiple of $2\pi$, where the integer
is equal to the half of the number of left and right crossings of the
  two oriented trajectories 
$\Gamma_a$ and $\Gamma_b$. This number is independent on the way in
  which the
  trajectories are projected on a plane and provides a
  well known alternative
  definition of the Gauss linking number.
In this way, also in the Coulomb gauge 
the condition that the Gauss linking number of $\Gamma_a$ and
$\Gamma_b$
should be equal
 to some value $m_{ab}$ is realized.
While abelian anyon field theories like those of Eq.~(\ref{CSaction})
may be significant in  quantum computing
\cite{abeliananyons}, it is rather
nonabelian statistics  that plays the main role  in this kind of
applications.
Despite its limitations, when the Gauss linking number is applied to a
$2s-$plat configuration, 
which cannot be destroyed because the $2s$ points of maxima and minima
are kept fixed, some of the nonabelian features of the system
are certainly captured. As a matter of fact, in the case of $2s-$plats
the capabilities of the Gauss linking number to distinguish the
changes of topology are  enhanced. The reason is the synergy
between the constraints imposed by the Gauss linking number
and those imposed by the fact that the
polymer system cannot 
escape the set of conformations allowed in a $2s-$plat.
 Indeed, since the end points of the subtrajectories
$\Gamma_{a,I_a}$ and  $\Gamma_{b,I_b}$ are fixed, also 
 the winding number between two different
subtrajectories is fixed. Due to the constraints imposed by the
linking number, allowed are only those
topology changes for which an amount of the winding angle of two
subtrajectories  is transferred in units of $2\pi$ to the winding
angle of another pair of subtrajectories. Moreover, the number
of subtrajectories is fixed to be equal to $2s$.
As a consequence, at least in the particular case of $2s-$plats, it is
possible to overcome the limitations of the Gauss linking number. As
mentioned in the Introduction,
if we start from an unlink consisting of a $4-$plat, the
system will never be able to attain the configuration of a Whitehead
link and viceversa, a $6-$plat Whitehead link cannot turn into a $4-$plat.
Moreover, in a forthcoming publication we will show how the present formalism
can be applied to include the treatment of links without the
limitations of the Gauss linking number and even to the case of
nontrivial knots.
This will pave the way to the treatment of polymer knots or links
constructed from tangles. Polymers of this kind are relevant in
biochemistry because nontrivial knot configurations appearing as a major 
pattern in DNA rings are mostly in the form of tangles \cite{sumners}.
\section{Acknowledgments}
F. Ferrari would like to thank E. Szuszkiewicz for pointing out
Ref.~\cite{collins} and inspiring the present work.
We wish to thank
heartily also M. Pyrka, V. G. Rostiashvili and T. A. Vilgis for fruitful
discussions. 
The simulations reported in this work were performed in part using the HPC
cluster HAL9000 of the Computing Centre of the Faculty of Mathematics
and Physics at the University of Szczecin. 
\begin{appendix}
\section{The length $L$ of a directed polymer as a function of the height}\label{appendixA}
In this Appendix we consider the partition function
\begin{equation}
{\cal Z}=\int\!{\cal D}\boldsymbol r(z)e^{-S}
\end{equation}
where $S$ is the action of the free open
polymer, whose trajectory $\Gamma$
is parametrized by means of the height $z$ defined 
in some interval $[\tau_0,\tau_1]$:
\begin{equation}
S=g
\int_{\tau_0}^{\tau_1}dz\left|\frac{d\boldsymbol r}{dz}\right|^2
\end{equation}
We want now to determine how the total length  of the curve $\Gamma$
depends on the constant parameter $g$. 
To understand what we mean by that, let us consider the standard
case of an ideal chain whose trajectory is
parametrized with the help of
the arc-length $\sigma$.
We denote with $a$ the average statistical length (Kuhn length)
of the $N$ segments  composing the polymer.
In the limit of large $N$ and small $a$ such that the product
$Na$ is constant, the total
length $L$ of the polymer 
satisfies the relation
\begin{equation}
L=Na \label{standlenght}
\end{equation}
We wish to obtain a similar identity connecting $L$ with $N$ and $g$
in the present situation, which is
 somewhat different.
To this purpose, we first dicretize the interval of integration
$[\tau_0,\tau_1]$ splitting it into $N$ small segments of length: 
\begin{equation}
\Delta z=\frac{\tau_1-\tau_0}{N}
\end{equation}
As a consequence, we may approximate the action as follows:
\begin{equation}
S\sim g\sum_{w=1}^{N}\left|\frac{\Delta \boldsymbol r_{w}}{\Delta
  z}\right|^{2}\Delta z
\end{equation}
where the symbol $\Delta \boldsymbol r_{w}$ means
\begin{equation}
\Delta \boldsymbol r_{w}=\boldsymbol r_{w+1}-\boldsymbol r_{w}
\end{equation}
and
\begin{equation}
\boldsymbol r_{w}=\boldsymbol r(\tau_0+w\Delta z)
\end{equation}
The discretized partition function becomes thus the partition function
of a random chain composed by $N$ segments: 
\begin{equation}
{\cal Z}_{disc}=\int\!\prod_{w=1}^{N}d\boldsymbol
r_{w}e^{-\sum\limits_{w=1}^{N}g\frac{|\Delta \boldsymbol
    r_{w}|^{2}}{\Delta z}}  \label{zdiscr} 
\end{equation}
Using simple trigonometric arguments
it is easy to realize that
the  length of each segment is:
\begin{equation}
\Delta L=\sqrt{|\Delta \boldsymbol r_{w}|^{2}+(\Delta z)^{2}}
\end{equation}
This is of course an average length,
dictated by the fact that, from Eq.~\ceq{zdiscr},
the values of $|\Delta \boldsymbol{r}_{w}|$ should be gaussianly
distributed around the point: 
\begin{equation}
|\Delta \boldsymbol{r}_{w}|^{2}=\frac{\Delta z}{g}
\end{equation}
In the limit $\Delta z \rightarrow 0$,
 the distribution of length of $\Delta \boldsymbol{r}_{w}$ becomes the Dirac
 $\delta$-function: 
\begin{equation}
\lim_{\Delta z\rightarrow 0}\frac{1}{2}\sqrt{\frac{g}{\Delta
    z}}e^{-g|\Delta \boldsymbol{r}_{w}|^{2}/\Delta z}\sim \delta 
 \left(|\Delta \boldsymbol{r}_{w}|-\sqrt \frac{\Delta z}{g}\right)
\end{equation}
If $N$ is large enough, we can therefore conclude that the total
length of the chain $\Gamma$ is: 
\begin{equation}
L\sim N\Delta L=N\sqrt{\frac{\Delta z}{g}+(\Delta z)^{2}}
\end{equation}
Since $N\Delta z=\tau_1-\tau_0$, we get:
\begin{equation}
L^{2}=|\tau_1-\tau_0|^{2}+\frac{N(\tau_1-\tau_0)}{g} \label{pollength} 
\end{equation}
In the limit $N\rightarrow \infty$, while keeping the ratio
$\frac{N}{g}$ finite, Eq.~(\ref{pollength}) 
 becomes the desired relation between the length of $\Gamma$ and
$g$ which replaces Eq.~\ceq{standlenght}. 

\section{The expression of the Gauss linking invariant in the Coulomb gauge}
\label{appendixB}
\begin{figure}[bpht]
\centering
\includegraphics[scale=.5]{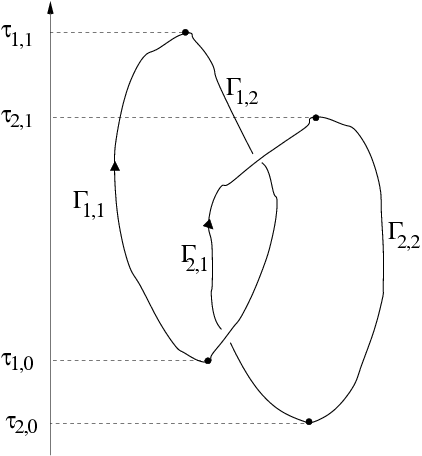}\\
\caption{Example of configuration of a $4-$plat.}\label{fig5}
\end{figure}

To fix the ideas, we will study here the particular case of a $4-$plat.
In the partition function \ceq{ppm} we isolate only the terms in which
the BF fields appear, 
because the other contributions are not connected to topological
constraints and thus are not relevant. 
As a consequence, we have just to compute the following partition function:
\begin{equation}
{\cal Z}_{BF,CG}(\lambda)=\int\!{\cal D}B_{\mu} {\cal
  D}C_{\mu}e^{-iS_{BF,CG}-S_{top}} \label{ZCSaux} 
\end{equation}
where the BF action in the Coulomb gauge $S_{BF,CG}$ has
been already defined in Eq.~\ceq{CSCG} 
and $S_{top}$ has been given in Eq.~\ceq{stop}.
In the case of a $4-$plat, $S_{top}$ becomes:
\begin{eqnarray}
S_{top}&=&i\lambda
\int_{\tau_{1,0}}^{\tau_{1,1}}dt\left[\frac{dx_{1,1}^{\mu}(t)}{dt}
B_{\mu}(\boldsymbol{r}_{1,1}(t),t)- 
\frac{dx_{1,2}^{\mu}(t)}{dt}B_{\mu}(\boldsymbol{r}_{1,2}(t),t) \right]\nonumber \\
&+&\frac{i\kappa}{8\pi^2}\int_{\tau_{2,0}}^{\tau_{2,1}}dt
\left[\frac{dx_{2,1}^{\mu}(t)}{dt}C_{\mu}(\boldsymbol{r}_{2,1}(t),t) 
-\frac{dx_{2,2}^{\mu}(t)}{dt}C_{\mu}(\boldsymbol{r}_{2,2}(t),t)\right]
\end{eqnarray}
where we recall that $x_{a,I}^{\mu}(t)=(\boldsymbol{r}_{a,I}(t),t)$,
$a=1,2$, $I=1,2$. For simplicity of the notation, in this Appendix we
use $\lambda$ 
instead of $\lambda_{12}$.
Using the Chern-Simons propagator of Eqs.~\ceq{propone}-\ceq{proptwo},
it is easy to evaluate the path integral over 
the gauge fields in Eq.~\ceq{ZCSaux}.
The result, after two simple Gaussian integrations, is:
\begin{equation}
Z_{BF,CG}(\lambda)=\exp{\left\lbrace
  \frac{i\lambda}{2\pi}\sum_{I,J=1}^2(-1)^{I+J-2}
\epsilon_{ij} 
\int_{\tau_0}^{\tau_1}d(x_{1,I}^{i}(t)-x_{2,J}^{i}(t))
\frac{(x_{1,I}^{j}(t)-x_{2,J}^{j}(t)) 
}{\left|\boldsymbol{r}_{1,I}(t)-\boldsymbol{r}_{2,J}(t)
\right|^{2}}\right\rbrace} \label{resZCS} 
\end{equation}
In the above equation we have put for simplicity:
\begin{eqnarray}
\tau_0&=&\max[\tau_{1,0},\tau_{2,0}]
\nonumber\\
\tau_{1}&=&\min[\tau_{1,1},\tau_{2,1}]
\end{eqnarray}
For instance, if the polymer configurations are as in Fig.~\ref{fig5}, we
have that $\tau_0=\tau_{1,0}$ and $\tau_1=\tau_{2,1}$.
Moreover, we remember that in our notation 
$\boldsymbol{r}_{a,I}(t)=(x_{a,I}^{1}(t),x_{a,I}^{2}(t))$.
Apparently, the elements 
of the trajectories $\Gamma_{1}$ and $\Gamma_{2}$ which lie below
$\tau_0$ and above $\tau_1$ 
do not take the part in the topological interactions. Thus is due to
the presence of the Dirac $\delta$-function $\delta(t-t^{\prime})$ 
in the components of the Chern-Simons propagator \ceq{propone}-\ceq{proptwo}.
However, we will see later that also the contributions
of these missing parts are present in 
the expression of $Z_{BF,CG}(\lambda)$.
In order to proceed, we notice that the exponent of the right hand
side of Eq.~\ceq{resZCS} consists 
 in a sum of integrals over the time $t$ of the kind:
\begin{equation}
D_{1,I;2,J}(\tau_1)-D_{1,I;2,J}(\tau_0)=
\epsilon_{ij}\int_{\tau_0}^{\tau_1}d\left(x_{1,I}^{i}(t)-x_{2,J}^{i}(t)
\right)\frac{(x_{1,I}^{j}(t)-x_{2,J}^{j}(t))
}{\left|\boldsymbol{r}_{1,I}(t)-\boldsymbol{r}_{2,J}(t)\right|^{2}}
\end{equation}
The above integrals can be computed exactly. It is in fact
well known that the function $D_{1,I;2,J}(t)$ is the winding angle of the
vector $\boldsymbol r_{1,I}(t)-\boldsymbol r_{2,J}(t)$ at time $t$: 
\begin{equation}
D_{1,I;2,J}(t)=\arctan \left(\frac{x_{1,I}^{1}(t)-x_{2,J}^{1}(t)}{x_{1,I}^{2}(t)-x_{2,J}^{2}(t)}\right)\label{dfunction}
\end{equation}
Thus, the quantity $D_{1,I;2,J}(\tau_1)-D_{1,I;2,J}(\tau_0)$ is a
difference of winding angles which measures  
how many times the trajectory $\Gamma_{1,I}$ turns around the trajectory $\Gamma_{2,J}$ in the slice of time $\tau_0\leq t\leq \tau_1$.
At this point, without any loss of generality,
we suppose that the configurations of the curves
$\Gamma_1$ and $\Gamma_2$ 
are such that the maxima and minima $\tau_{a,I}$ are ordered as follows:
\begin{equation}
\tau_{2,0}<\tau_{1,0}<\tau_{2,1}<\tau_{1,1}
\end{equation}
As example of loop configurations that respect this ordering is
given in Fig.~\ref{fig5}. 
As a consequence, we have:
\begin{equation}
\tau_0=\tau_{1,0} \qquad \mbox{and}\qquad \tau_1=\tau_{2,1}
\end{equation}
Now we notice that the logarithm of the gauge partition function ${\cal
  Z}_{BF,CG}(\lambda)$ in Eq.~\ceq{resZCS} contains a sum of
differences of the winding angles defined in Eq.~(\ref{dfunction}): 
\begin{eqnarray}
\frac{2\pi\log{{\cal
      Z}_{BF,CG}}(\lambda)}
{i\lambda}&=&\left[D_{1,1;2,1}(\tau_{2,1})-D_{1,1;2,1}(\tau_{1,0})+ 
D_{1,2;2,2}(\tau_{2,1})-D_{1,1;2,2}(\tau_{2,1})\right.\nonumber \\
&+&\left.D_{1,2;2,1}(\tau_{1,0})-D_{1,2;2,1}(\tau_{2,1})+
D_{1,1;2,2}(\tau_{1,0})-D_{1,2;2,2}(\tau_{1,0})\right] \label{logZCS}
\end{eqnarray}
Further, assuming that the curves $\Gamma_1$ and $\Gamma_2$ are oriented as in
Fig.~\ref{fig5}. 
if we start from the minimum point at $\tau_0=\tau_{1,0}$, we can
 isolate in the right hand side of Eq.~\ceq{logZCS} 
the following four contributions:
\begin{enumerate}
\item
In the time slice $\tau_{1,0}\leq t\leq \tau_{2,1}$
the angle which measures the winding of the trajectory $\Gamma_{1,1}$
around the trajectory $\Gamma_{2,1}$ is given by the difference
$D_{1,1;2,1}(\tau_{2,1})-D_{1,1;2,1}(\tau_{1,0})$. 
\item
In the region $\tau_{2,1}\leq t\leq \tau_{1,1}$ only the trajectory
$\Gamma_1$ continues to evolve, 
going first upwards with the subtrajectory $\Gamma_{1,1}$ and then downwards
with $\Gamma_{1,2}$.
After this evolution, the winding angle 
between the two trajectories $\Gamma_1$ and $\Gamma_2$ has changed by
the quantity $D_{1,2;2,2}(\tau_{2,1})-D_{1,1;2,2}(\tau_{2,1})$.
\item
Next, in the region $\tau_{2,1}\geq t\geq \tau_{1,0}$, the winding
angle which measures how many 
times the subtrajectory $\Gamma_{1,2}$ winds up around 
$\Gamma_{2,2}$ is given by the 
difference $D_{1,2;2,1}(\tau_{1,0})-D_{1,2;2,1}(\tau_{2,1})$.
\item
Finally, in the region $\tau_{1,0}\geq t\geq \tau_{2,0}$ only the
second trajectory $\Gamma_2$ continues 
to evolve, going first downwards with the curve $\Gamma_{2,2}$ and
then upwards with $\Gamma_{2,1}$. 
The net effect of this evolution is that the 
winding angle between
$\Gamma_1$ and $\Gamma_2$ changes
by the quantity $D_{1,1;2,2}(\tau_{1,0})-D_{1,2;2,2}(\tau_{1,0})$.
\end{enumerate}
It is thus clear that the right hand side of Eq.~\ceq{logZCS}, apart
from a proportionality factor $i\lambda$,
counts how many times the trajectory $\Gamma_{1}$ winds around the
trajectory $\Gamma_{2}$. If we wish to identify the quantity in the
right hand side of Eq.~\ceq{logZCS} with the Gauss linking number
$\chi(\Gamma_1,\Gamma_2)$, we should check
for consistency that it  takes only integer values as
the Gauss linking number does. 
Indeed, it is easy to see that, modulo $2\pi$, the
following identities are holding: 
\begin{eqnarray}
D_{1,1;2,1}(\tau_{2,1})&=&D_{1,1;2,2}(\tau_{2,1})\nonumber\\
D_{1,1;2,2}(\tau_{1,0})&=&D_{1,2;2,2}(\tau_{1,0}) \nonumber \\
D_{1,2;2,2}(\tau_{2,1})&=&D_{1,2;2,1}(\tau_{2,1})\nonumber\\
D_{1,1;2,1}(\tau_{1,0})&=&D_{1,2;2,1}(\tau_{1,0})
\end{eqnarray}
For example, the first of the above equalities states that the 
angle formed by the vector $\boldsymbol r_{1,1}-\boldsymbol r_{2,1}$
connecting the subtrajectories $\Gamma_{1,1}$ and $\Gamma_{2,1}$ at
the height $\tau_{2,1}$ is equal to the angle
formed by the vector $\boldsymbol r_{1,1}-\boldsymbol r_{2,2}$
connecting the subtrajectories $\Gamma_{1,1}$ and $\Gamma_{2,2}$ at
the same height. The reason of this identity is trivial: At that
height, the subtrajectories $\Gamma_{2,1}$ and $\Gamma_{2,2}$ are
connected together at the same point.
Applying the above relations to Eq.~\ceq{logZCS}, one may prove that:
\begin{equation}
\frac{2\pi\log{{\cal Z}_{BF,CG}}(\lambda)}{i\lambda}=0\qquad\qquad\mod 2\pi
\end{equation}
As a consequence, we can write:
\begin{equation}
{\cal Z}_{BF,CG}(\lambda)=e^{i\lambda\chi(\Gamma_1,\Gamma_2)} \label{Z}
\end{equation}
where $\chi(\Gamma_1,\Gamma_2)$ is the Gauss linking number.
Concluding, the above analysis shows that also in the Coulomb gauge the
BF fields in the polymer partition function
\ceq{ppm} fix the topological constraints \ceq{topconst} correctly, in
full consistency
with the results obtained in the covariant gauges. 
Of course this consistency was expected due to gauge invariance.
Yet, it is interesting that, using the Coulomb gauge, one may express
the Gauss 
linking number invariant in a way that is quite different from the usual
form given in Eq.~\ceq{ln}.
\end{appendix}

\end{document}